\documentclass[a4paper,fleqn,usenatbib]{mnras}

\usepackage{newtxtext,newtxmath}
\usepackage[T1]{fontenc}
\usepackage{ae,aecompl}

\usepackage{graphicx}	% Including figure files
\usepackage{amsmath}	% Advanced maths commands
\usepackage{amssymb}	% Extra maths symbols
\usepackage{wasysym}
\usepackage{float}
\newcommand\msun{\rm M$_{\odot}$}

\newcommand\review{}
\newcommand\reviewtwo{}
\title[Star cluster formation with photo-ionisation]{Star Cluster Formation in a Turbulent Molecular Cloud Self-Regulated by Photo-Ionisation Feedback}

\author[E. Gavagnin et al.]{
Elena Gavagnin$^{1}$,\thanks{E-mail: gavagnin@physik.uzh.ch}
Andreas Bleuler$^{1}$,
Joakim Rosdahl$^{2}$
and Romain Teyssier$^{1}$
\\
$^{1}$Institute for Computational Science, Centre for Theoretical Astrophysics and Cosmology,\\ \, Universit\"at Z\"urich, CH-8057, Z\"urich, Switzerland\\
$^{2}$Univ. Lyon, Univ. Lyon1, Ens de Lyon, CNRS, Centre de Recherche Astrophysique de Lyon UMR5574,\\ \, F-69230, Saint-Genis-Laval, France
}

\date{\today}

\pubyear{2017}

\begin{document}
\label{firstpage}
\pagerange{\pageref{firstpage}--\pageref{lastpage}}
\maketitle

\begin{abstract}
Most stars in the Galaxy are believed to be formed within star clusters from collapsing molecular clouds. 
However, the complete process of star formation, from the parent cloud to a gas-free star cluster, is still poorly understood. 
We perform radiation-hydrodynamical simulations of the collapse of a turbulent molecular cloud using the {\ttfamily RAMSES-RT} code. 
Stars are modelled using sink particles, from which we self-consistently follow the propagation of the ionising radiation. 
We study how different feedback models affect the gas expulsion from the cloud and how they shape the final properties of the emerging star cluster.
We find that the star formation efficiency is lower for stronger feedback models. Feedback also changes the high mass end of the stellar mass function.
Stronger feedback also allows the establishment of a lower density star cluster, which can maintain a virial or sub-virial state. 
In the absence of feedback, the star formation efficiency is very high, as well as the final stellar density. 
As a result, high energy close encounters make the cluster evaporate quickly. 
Other indicators, such as mass segregation, statistics of multiple systems and escaping stars confirm this picture. 
Observations of young star clusters are in best agreement with our strong feedback simulation. 
%add stuff
\end{abstract}


%%%%%%%%%%%%%%%%%%%%%%%%%%%%%%%%%%%%%%%%%%%%%%%%%%

%%%%%%%%%%%%%%%%% BODY OF PAPER %%%%%%%%%%%%%%%%%%

\section{Introduction}

Establishing a full and consistent theory of star cluster formation remains an open task for the scientific community. 
The most widely adopted view is that star clusters form from the collapse of giant molecular clouds. 
On a timescale of a few millions years, a cloud undergoes gravitational collapse and converts part of its gas into many dense molecular cores, 
each core leading to the formation of one or a few proto-stellar objects \citep[see][for a review]{KlessenSF, KrumholzSF}. 
These protostars can continue accreting material from their surroundings, and eventually become proper stellar, main sequence objects, 
whose stellar luminosity is high enough to inject considerable amounts of energy into their parent cloud. 
This stellar feedback modifies the properties of the cloud and the star formation process itself and as a result regulates the properties of the emerging star cluster, 
such as its dynamical state, the mass distribution and the fate of its stellar population. 

Understanding the impact of stellar feedback on the star cluster properties, and the transition from the initial turbulent GMC to the final gas-free association of stars 
(such as observed open clusters, embedded clusters or even globular clusters) is at the moment one of the most intriguing fields of research in astrophysics, 
mainly because of the numerous and complex physical processes at play during the entire history of the star cluster formation. 

A classic reference is the work of \cite{LadaLada}, which states that 90\% of stars are likely to form in star clusters . 
In \cite{LadaLada}, star clusters  are defined as groups of at least 35 stars and with a stellar mass density of at least 1~\msun $\rm pc^{-3}$. 
These numbers can be derived by requiring that the evaporation timescale of the star cluster is longer than 100~Myr.
A more recent study by \cite{Bressert10} revealed how the fraction of stars in the solar neighbourhood forming in clusters is strongly dependent on the adopted definition for star clusters , 
with values ranging between 45 and 90\%. They concluded that stars form within a broad and smooth distribution of surface densities, 
which is consistent with star formation proceeding hierarchically, within the turbulent, hierarchical structure of the parent molecular cloud, 
where denser regions are systematically embedded in less dense regions \citep{Elmegreen06, Bastian07}.

Defining what is a truly bound cluster or an unbound stellar association is indeed not straightforward, especially when the system is young. 
It is only after these stellar structures have dynamically evolved, that they are easier to distinguish from their environment. 
The identification of the fraction of stars residing within these older stellar systems is more reliable, and is observed to be around 10-30\% \citep{Miller78,Adamo11}.
{\review \cite{Kruijssen2012} shows that the cluster-formation efficiency varies from 1-70\% depending on the galactic gas surface densities at which the cluster forms.}

It is also very important to establish what is the fraction of stars which formed in star clusters  but do not reside there anymore today. 
This is usually referred as star clusters  {\it infant mortality}, outlining the fact that, when we compare the fraction of stars in young, embedded star clusters  
with the fraction of stars in older, open clusters, most of the clusters seems to have been disrupted during this transition from embedded to exposed \citep{LadaLada}. 
Note that this interpretation assumes that the fraction of stars in star clusters  is the rather old one presented in \citep{LadaLada}. 

The commonly adopted picture for the cause of this infant mortality is the fast expulsion of the initial gas, leading to the rapid expansion and disruption of the star cluster.
Only clusters with a star formation efficiency (SFE, i.e. the fraction of gas converted into stars) higher than 30\% are believed to survive the gas removal and stay bound \citep{Hills80,Lada84,Bastian06}. 
Yet, the star formation efficiency is not the only parameter that can decide whether a star cluster will survive gas expulsion.
Two other important factors are: 1-the timescale of gas removal and 2-the actual dynamical state of the star cluster right before expulsion. 
Regarding the first point, it has been shown for example that systems with star formation efficiency as low as 10\% can remain bound, 
as long as the gas is removed slowly and adiabatically \citep{Baumgardt07}. 
The second factor has been pointed out by \cite{Goodwin09}, showing a strong dependence of the star cluster mass loss (hence survival) on the virial ratio of the emerging star cluster. 
%Indeed, if these are not in virial equilibrium immediately before the onset of gas expulsion, the effective star formation efficiency (SFE) does not correspond anymore to the true final SFE. 
%In practice, this means that i
Indeed, if the system is sub-virial before gas is expelled, it can survive even with SFE lower than 30\%. 
Conversely, an initially super-virial system, even with a SFE as high as  50\%, will be at edge of survivability \citep{Goodwin09}.

{\review \cite{Kruijssen2012} questions the importance of gas expulsion in determining the fate of the star cluster and justifies the observed poor number of bound clusters as direct result of the star formation process. According to the author, most of the natal cloud is characterised by low SFE and will therefore form dispersed structure, while only the few sites of high SFE will give birth to bound star clusters.}

The SFE within star forming molecular clouds is poorly understood from theoretical grounds.
Simple models based only on self-gravitating turbulence predict a very high SFE, higher than 90\%, meaning that 
star formation occurs during one free-fall time of the parent cloud, in contradiction with observational constraints \citep{Padoan14}. 

Stellar feedback has been invoked to reduce the SFE by terminating star formation in giant molecular clouds \citep[see the review by][and references therein]{DaleReview}.  
Stellar feedback is a broad term that refers to the injection of mass, momentum and energy by stars and protostars into the star forming gas itself. 
The different mechanisms of stellar feedback are photoionisation from massive main sequence stars, infrared and optical radiation from accreting protostars, 
radiation pressure associated to these various types of radiation, proto-stellar jets, stellar winds from main sequence or post-main sequence stars, supernovae explosions. 
Although all these ingredients are likely to play an important role in regulating the star formation efficiency and in setting the properties of the emerging star clusters,  
they act on different spatial and temporal scales, and are associated with stars of different masses. {\review During the first Myrs of a star cluster life, before the first OB stars form, feedback modes from pre-main sequence stars play a significant role. These include jets, deuterium-burning and accretion feedback. Pre-main sequence feedback is generally not effective on large-scale and does not drive the process of gas clearing, however it has been shown to be able to sustain turbulence and reduce the conversion rate of gas into stars \citep{Krumholz2012, Federrath2015}. Moreover radiation focusing in the direction of outflow cavities prevents the formation of radiation pressure-supported gas bubbles, diminishing the radiative heating and outward radiation force exerted on the infalling cloud gas \citep{Cunningham2011}, resulting in higher mass accretion onto the protostar. Disk fragmentation is also suppressed as a result of thermal feedback from protostars \citep{Offner2009}, affecting the multiplicity of stellar systems.}%you didn't discuss directly what would have changed in your simulations if these effects are included. 

On the observational side, several surveys can be used to cast light on the star cluster formation process. 
The MYSTiX survey \citep{Mystix}, for example, is targeting massive star forming regions and has revealed that star clusters are frequently divided into sub-clusters \citep{Kuhn15}. 
We now have evidence that these sub-clusters  are expanding or merging, with clear signs of ongoing dynamical relaxation. For example, we observe mass segregation (see Section \ref{massseg} for a definition) down to 1.5 \msun \citep{Kuhn15}.
Similarly, \cite{DaRio14} have studied the morphology and the dynamical state of the Orion Nebula Cluster. They concluded that the core appears rounder and smoother than the outskirts, 
which is consistent with ongoing dynamical processing. 

The Gaia-ESO Survey \citep{GaiaEso} has recently discovered several kinematically distinct populations in the young star cluster Gamma Velorum, 
surrounding the $\gamma^2$ Velorum binary in the Vela OB2 association.
%and one in NGC2547 \citep{Sacco15}. 
%The main hypothesis on the origin of the distinct population in  NGC2547 are that it could have originally be part of a cluster around $\gamma^2$ Velorum, which expanded due to gas expulsion or it could have formed in a less dense environment \citep{Sacco15}. 
According to \cite{Jeffries14}, the first component of Gamma Velorum is a bound remnant of an initially larger cluster, formed in a dense region of the Vela OB2 association, 
that has been partially disrupted by gas expulsion. The second component consists of a scattered population of unbound stars born later 
(as indicated by lithium depletion) in less dense regions. The gas surrounding this second population was probably evaporated by the radiation coming from the first one, 
quenching the star formation episode quite abruptly.

In general, very young star clusters, sometimes still embedded in their parent gas cloud, are ideal laboratories to study the effect and phenomenology of stellar feedback and gas expulsion. 
In the Milky way, the so-called ``starburst star clusters'' (e.g. NGC 3603 YC, Quintuplet, Arches, Westerlund 1 and 2) 
represent the youngest (< 5 Myr) and more actively star forming clusters \citep{Brandner08}. 
NGC 3603 YC, for example, is only $\rm \sim$ 1 Myr old, and is surrounded by glowing interstellar gas and obscuring dust \citep{Rollig11}. 
The Arches, the second youngest with an age of $\sim$ 2.5 Myr, is already free of any gas in its centre \citep{Stolte03} with a clear X-ray signature of hot outflowing gas \citep{Yusef02}. 
These newborn star clusters are characterised by the presence of strongly UV-radiation from O and B stars that ionises the nebula and disperses the gas \citep{Crowther10, McLeod16}. 

%The kinematics of the two populations in Gamma Velorum is also compatible with a marging scenario in which the two population were originated from two different molecular cores, being part of the same molecular cloud \citep{Mapelli15}. 

On the theoretical side, the challenge of modelling star clusters is due to the lack of a complete theory of star formation. This is an inherently multi-scale, multi-physics problem,
with a central role played by feedback mechanisms. We point to the reviews by \cite{DaleReview} and \cite{KrumholzReview} for a detailed presentation of the problem. 
Here we present only a few selected earlier studies, relevant for our work which focuses specifically on the star cluster formation process.

{\review \cite{Walch12} and \cite {Walch2013} modelled fractal clouds by means of 3D smoothed particle hydrodynamics simulations and explored the effect of a ionising O-star at the centre of a $\rm 10^4$ \msun} {\review cloud. They found that some global properties, such as the total outflow rate, the distribution of gas into high- and low-density and the injected kinetic energy are all independent of the fractal dimension, while the statistical properties of the triggered star formation events and the shell morphology both appear to correlate with the cloud fractal dimension.}

\cite{Fujii15}, \cite{Fujii15b} and \cite{Fujii16} used direct N-body simulations, 
starting from initial conditions drawn from the results of previous smoothed particle
hydrodynamics (SPH) simulations of turbulent molecular clouds. 
Because the adopted SPH resolution was relatively low ($\sim~0.1$~pc), the authors could not resolve the formation of individual stars, 
but could still capture the clumpy structure of the gas.  
After one free-fall time of the initial gas cloud, they stopped the hydro simulation and replaced dense enough gas particles with stellar particles, 
assuming a star formation efficiency (or gas to star conversion factor) depending on the local gas density. 
The remaining gas particles were removed instantaneously and the stellar particles dynamics was integrated further in time using a direct N-body code.
They derived that the initial properties of the parent cloud (mass, density) determine the characteristics of the emerging cluster, whether it will become an association, an open cluster or a dense massive one. Moreover, to form massive clusters, they claimed that a local star formation efficiency >50\% is needed.

Using a more elaborate methodology, \cite{DaleBon11}, \cite{DaleBon12}, \cite{Dale12a}, \cite{Dale12b}, \cite{Dale13a} and \cite{Dale13b} studied in a series of papers 
the effects of photo-ionisation feedback on embedded clusters and its disruptive impact on clouds of different masses (from 10$^4$ to 10$^6$~M$_\odot$) 
and sizes (from 2 to 220~pc), either initially bound or unbound.
In \cite{Dale14}, the authors added stellar winds to photoionisation feedback and studied how the overall star formation efficiency, the average star formation rate (SFR) 
and the fraction of unbound gas varied with the initial cloud properties. 
Their methodology was based on SPH simulations of turbulent molecular clouds, with an initial shallow Gaussian density profile.
The velocity field was initialised as a turbulent, divergence-free Gaussian random field, with a power spectrum to $P(k) \propto k^{-4}$ consistent with isothermal supersonic turbulence. 
The cloud was evolved using self-gravity and cooling, and star formation was modelled using sink particles. The mass and spatial resolution was also relatively low, with {\reviewtwo $10^6$ particles per cloud, but using $100$ neighbours in the smoothing kernel, so only $10^4$ independent resolution elements \citep{Dale2007}}. Radiative transfer of the photo-ionising photons was performed using a Str\"omgren sphere filling technique \citep[see][for details]{Dale07}.
Using the same set of simulations, \cite{Dale15} focused on the properties of the stellar populations of the star clusters formed.  
They found that the star formation efficiency is lowered by the presence of feedback, however they stressed how the disruptive effect of feedback depends on the cloud properties, especially the escape velocity. Natal gas from massive clouds with elevated escape velocities is expelled only in minimal part. Winds are found to have little impact on the dynamics of gas compared to ionising feedback. Moreover, in these simulations the number of stars unbound by feedback is very modest and is not related to the fraction of gas expelled.

Along the same lines as in \cite{Fujii15}, the longer term evolution of these star clusters was finally investigated in another series of paper by \cite{Parker13, Parker15, Parker15b}.
They concluded that clusters formed in simulations with feedback tend to remain sub-structured longer than in the non-feedback cases. Moreover, at the end of the pure N-body evolution, the authors found that simulations with feedback contain fewer bound stars than in the control run. In terms of mass segregation, they do not provide a unique conclusion, because different analysis return contrasting results.

More recently, several papers have addressed the problem of star cluster formation from a realistic, gaseous, turbulent environment using grid-based simulation techniques.
Using the {\ttfamily RAMSES} code, \cite{Lee:2016gz} studied the conditions required in the parent cloud to obtain a bound star cluster.
The authors aimed to examine the properties of the gaseous proto-cluster born from the collapse of a $10^4$ \msun molecular cloud. To achieve this they performed magnetohydrodynamics simulations, without stellar feedback and varying the initial level of turbulent support. Prestellar cores were followed using the same sink particles algorithm adopted in our work. The typical mass of a sink was $10$ \msun. The proto-cluster turned out to be in virial equilibrium, with turbulence and rotation supporting the collapse. The virial status and size of the proto-cluster were considered to be directly imprinted by the parent cloud, therefore they concluded that the study of the gaseous proto-cluster phase is a fundamental step in the context of stellar cluster formation.

Using the FLASH code, coupled to a ray tracing code, \cite{2016MNRAS.461.2953H} studied the effect of various cloud initial conditions, then subjected to the ionising radiation of massive stars,
on the final properties of the star cluster system. 
This study focused on giant $10^6 M_\odot$ molecular clouds, with different initial virial parameters ($\alpha$), ranging from bound ($\alpha=0.5$)  to unbound ($\alpha=5$). 
The main goal was to study how feedback and the virial status affect the formation of star clusters and subsequent evolution of the cloud. In this case sink particles represented single star clusters and star formation within each cluster is implemented with a subgrid model, by randomly sampling the IMF. Their conclusion was that the initial virial parameter strongly influences the SFE, with more bound clouds having higher efficiency, while radiative feedback did not play a major role, lowering the previous values only by few percent.  
They also found that the number of star clusters formed depends on the boundedness of the cloud: the more bound the cloud, the fewer the star clusters. Moreover, the clusters from unbound clouds were gas poorer and star richer than the ones formed from bound clouds.

In this work, we model the collapse of a $\sim 2.5 \times 10^4$~M$_\odot$ turbulent cloud with photo-ionisation feedback from massive stars at extremely high resolution (smallest cell size $\sim$ 500 AU), 
and study how the star cluster forms and emerges from its parent cloud. Our radiative transfer technique is based on the moment method with the M1 closure \citep{2013MNRAS.436.2188R}
and allows to model an arbitrary number of photon sources, much faster than traditional ray tracing schemes.   
We consider two different feedback scenarios (strong and weak) and a reference simulation without any feedback. 
We subsequently analyse how the different feedback scenarios affect the properties of our new born star clusters, using various observables related to the stellar mass function, 
its spatial distribution, the mass segregation, the distribution of escaping stars and the stellar multiplicity function.

The paper is organised as follows: in Section~\ref{methods}, we describe the numerical methods we have used for our simulations. 
In Section~\ref{results}, we analyse the properties of the star clusters we have obtained, and finally, in Section~\ref{discussions}, we discuss our findings in light of previous studies, both theoretical and observational. 

\section{Numerical Methods}\label{methods}

We now describe in details the numerical techniques we use to model the collapse of a turbulent molecular cloud and the formation of massive stars,
following the effects of ionising radiation on the cloud itself.

\subsection{Initial Conditions}

We first perform a decaying turbulence simulation in a periodic box sampled with $1024^3$ cells. This simulation is initialised with a uniform gas density $\rho_0=1$ (in arbitrary units) 
and a Gaussian random velocity field with a power spectrum $P(k) \propto k^{-4}$, where $k$ is the wavenumber. $P(k)$ is normalised so that the 3D velocity dispersion in the full box was set to $\sigma_{\rm 3D} = {\cal M} c_s$, 
where the sound speed is $c_s=1$ in arbitrary units and the initial Mach number is set to ${\cal M}=20$. 
After one turbulence crossing time, $t_{\rm turb} = L / \sigma_{\rm 3D}$ (where the box size was also set to 1 in arbitrary units), the kinetic energy has decayed by a factor of two, 
and the actual Mach number by a factor of $\sqrt{2}$. At that time, the turbulence is fully developed, with density fluctuations following a clear log-normal distribution function 
and the variance in $\log \rho$ reaching its peak value. 
{\review \cite{Krumholz2012} found that whether turbulence is initially fully developed or not has significant impact on the results.}

We then use this final snapshot as a template for the initial turbulent cloud. We first set up the physical scales of our problem. 
The cloud is considered to be fully composed of molecular gas Hydrogen with temperature $T_0=10$~K and isothermal sound speed $c_s = 0.2$~km/s. 
The mean density in the box is set to $n_H=10^3$~H/cc and the periodic box length to 20~pc.
We carve out of the periodic box a sphere of radius 5~pc, centred on a large filament resulting from a large compressive mode. 
As a result, the mean density in the spherical cloud is larger than the mean density in the original box,
and the Mach number in the cloud is smaller than in the original box (by another factor of $\sqrt{2}$) with ${\cal M} \simeq10$.
The final cloud properties are the following: radius $R=5$~pc, mass $M \simeq 2.5\times10^4$~M$_\odot$ and velocity dispersion $\sigma_{\rm 3D} \simeq 2~$km/s.
Note that, because we have adopted a velocity dispersion at the low end of values found in observations of clouds of a similar size, our cloud virial parameter 
\begin{equation}
\alpha_{\rm vir} = \frac{5 \sigma^2_{\rm 3D} R}{3 G M} \simeq 0.3,
\end{equation}
is small enough to ensure a fast collapse, i.e. the free-fall time is $\sim$ 1 Myr.
The simulations are then run to t=2Myr.

{\reviewtwo 
%The initial value for the virial parameter was chosen so that the collapse proceeded fast and we could perform a first study on the effect of different photoionisation regimes. 
Such a choice for the virial parameter was meant to explore the stabilising effect of feedback. 
We chose, in fact, an extreme situation to investigate the action range of photoionisation even in very bound and dense environments, characterised by a high degree of dynamical interactions and escaping stars. 
Moreover, cloud disruption driven by large scale turbulence (see works by Dale et al.) is not effective for our cloud.
We intend to relax such an extreme condition in a follow-up paper.}

\subsection{Refinement strategy}

Our initial coarse grid corresponds to a minimum refinement level $\ell_{\rm min}=10$ with cell size $\Delta x_{\rm max}\simeq0.02$~pc, which allows us to resolve our sonic scale $l_s \simeq 0.08$~pc, i.e.
the scale at which our scale-dependent 3D velocity dispersion is equal to the sound speed. During the course of the simulation, we refine this initial grid level using a quasi-Lagrangian
refinement criterion. Our maximum resolution is fixed to our maximum refinement level $\ell_{\rm max}=13$, which corresponds to a minimum cell size of $\Delta x_{\rm min}\simeq 500$~AU.
Assuming for the isothermal sound speed $c_s = 0.2$~km/s, and requiring for the Jeans length 
\begin{equation}
\lambda_{\rm J}=c_s \sqrt{\frac{\pi}{G \rho}} > 4 \Delta x_{\rm min},
\end{equation} 
this gives us the constraint that $\rho < \rho_{\rm J} \simeq 2 \times 10^{-17}$~g/cc. This maximum density corresponds also to a Jeans mass
\begin{equation}
m_{\rm J} = \frac{4 \pi}{3}  \rho_{\rm J} \left( \frac{\lambda_{\rm J}}{2}\right)^3 \simeq 0.14~M_\odot.
\end{equation} 
We require to resolve this Jeans mass with at least 64 resolution elements, which gives us a mass resolution of $m_{\rm res} \simeq 2 \times 10^{-3}~M_\odot$.
Our refinement strategy is thus the following: if a cell has accumulated a gas mass larger than $m_{\rm res}$, then it is refined individually into 8 new children cells, up to the maximum refinement level.
Note that with our adopted initial coarse level and our quasi-Lagrangian strategy, we also automatically satisfy the additional criterion 
that the Jeans length is always refined by at least 4 cells for any gas density smaller than $\rho_{\rm J}$. 

\subsection{Sink Particles}

When the gas density exceeds $\rho_{\rm J}$, we violate our requirement to always resolve the Jeans length with 4 cells and the Jeans mass with 64 resolution elements.
Therefore we adopt this criterion to form sink particles, using the technique developed in \cite{2014MNRAS.445.4015B}. We first detect density peaks in our 3D density field using the 
PHEW clump finder \citep{2015ComAC...2....5B}. The density threshold is set to $\rho_{\rm threshold}=2\times 10^{-18}$~g/cc, 
or 10\% of the Jeans density. After we have identified 
a discrete set of peak patches delimited by either the isosurface at the density threshold or the saddle surface with a neighbouring peak patch, we draw a sphere, 4 cell size in radius,
around the density maximum. If the density at the maximum exceeds the Jeans density, if the sphere is contracting and if its virial parameter is less than 1,   
we form a sink with a seed mass equal to $m_{\rm J} \simeq 0.14~M_\odot$ \citep[see][for details]{2014MNRAS.445.4015B}. In our simulations one sink corresponds to a single star.

The sink particle is then treated like a point mass. We follow the sink particles dynamics by a leap-frog, direct N-body integrator, using a softened $1/r^2$ acceleration (with softening length $0.5 \Delta x_{\rm min}$) between sinks, and also 
between the sinks and the gas. 
Only the self-gravity of the gas is based on the grid-based Poisson solver in {\ttfamily RAMSES}. Gas accretion onto the sink particles is modelled through what is described as ``flux accretion" in  \cite{2014MNRAS.445.4015B}. 
\subsection{Radiative Processes}

In this paper, we model the emission and the propagation of ionising, ultra-violet (UV) radiation, together with associated heating and cooling processes.
We used the {\ttfamily RAMSES-RT} radiative transfer module developed by \cite{2013MNRAS.436.2188R}, using one photon group, with energies between 
13.6~eV and 24.6~eV.%, between 24.6~eV and 54.4~eV and above 54.4~eV. 
We do not account for photon energies below 13.6~eV, namely optical and infrared radiation,
as the scope of the paper is to study the effect of photo-ionisation heating on the molecular cloud. We will study these other sources of radiation in a follow-up paper.
Details in the adopted photo-absorption cross section, chemistry and cooling processes are available in \cite{2013MNRAS.436.2188R}.
Metal cooling prescriptions are based on \cite{Sutherland1993} for temperatures above $10^4$ K and on \cite{Rosen1995} for metal fine-structure cooling below $10^4$ K.
{\review We extended the cooling function by \cite{Rosen1995} down to 10 K, to account for CO and fine structure cooling due to CII, OI, according to prescriptions of \cite{Dalgarno1972}.}
Following \cite{Geen2015a,Geen2016}, the photon group energy and cross-section are derived sampling the blackbody spectral energy distribution of a 20 \msun \, star.
The frequency-dependent ionisation cross sections are taken from \cite{Verner1996} and \cite{Hui1997}.
A reduced speed of light of $10^{-4} c$ is used. This is done to improve the efficiency of our simulations, since the speed of light affects the tilmestep calculation, through the Courant factor. 

\begin{figure}
\includegraphics[width=\columnwidth]{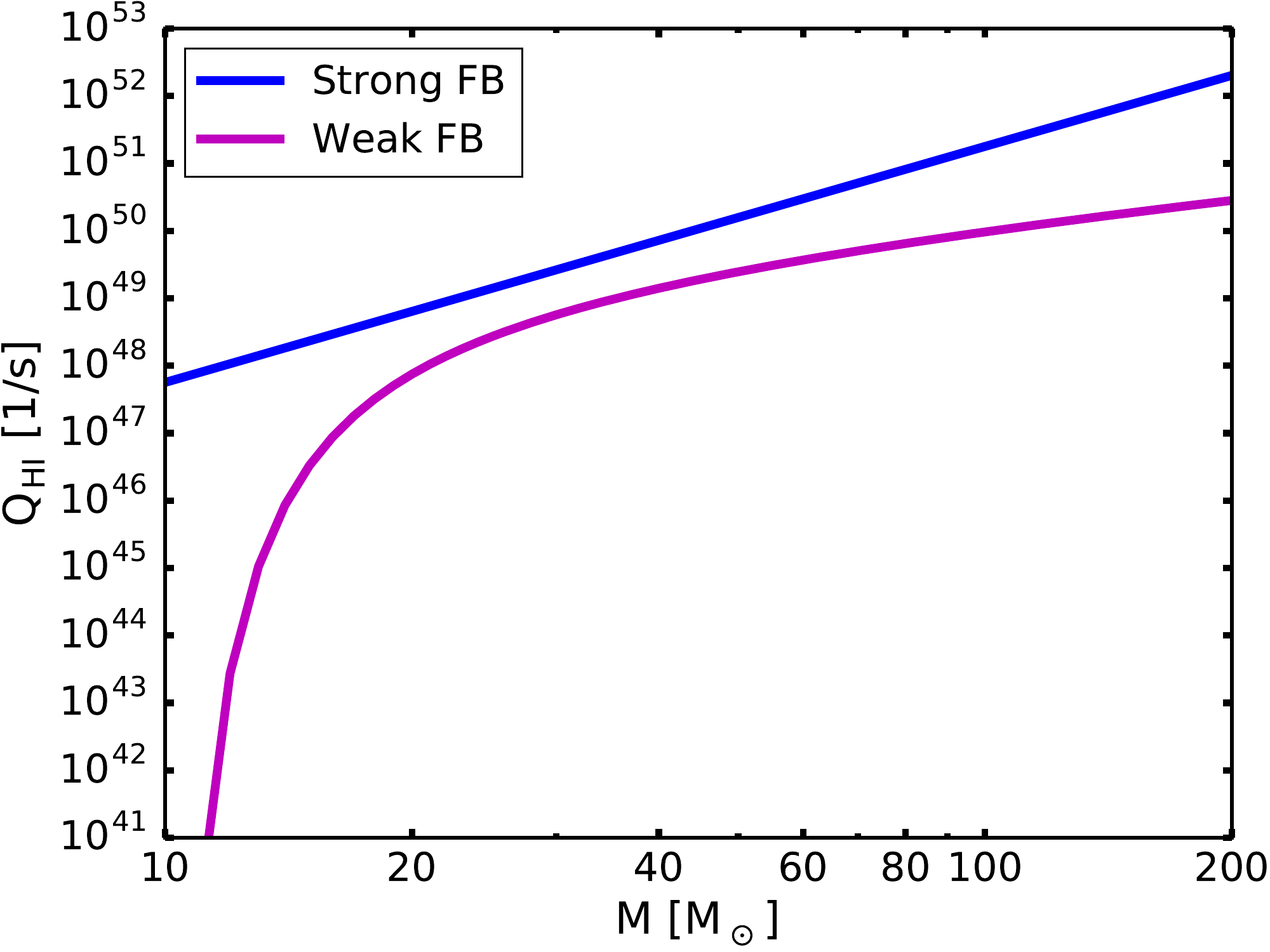}
\caption{Number of emitted ionising photons per second as a function of stellar mass.}
\label{Q_H}
\end{figure}

The UV radiation emitted by the sink particles is modelled using the following simple strategy.
We implemented two feedback regimes, namely {\it strong} and {\it weak}. 
For the strong feedback case, we basically consider all the energy emitted from the sink/star (even optical and infrared) as ionising radiation. To derive the energy associated with every sink we assume a power-law luminosity-mass relation, $\rm L~=~L_\odot (M/M_\odot)^{3.5}$, where $L_\odot$ and $M_\odot$ are the solar luminosity and solar mass, respectively .
{\review The number of photons emitted per second, $\rm Q_{HI}$, was then obtained by dividing this luminosity by a mean value of photon energy in the ionisation range (13.6eV-24.6 eV).}
 %The number of emitted ionising photons per second, $\rm Q_{HI}$, is then simply L/13.6~eV. 
For the weak feedback case, we computed an analytical fit of photon emission rates presented in \cite{Sternberg03}, obtained through radiation-driven wind atmosphere models of OB stars.
We derived the following analytic expression of the number of emitted ionising photons per second as a function of the stellar mass:
\begin{align}
\rm \log [Q_{HI} (M)] = \rm 48.65+\log(M/M_\odot) - \frac{2.4}{\log(M/M_\odot-8)^{1.9}} \, .
\end{align} 
This formula was applied to calculate emission rates for all sinks with $M > 10 M_\odot$. For stars with lower mass we assume there are no ionising photons.
{\review Figure \ref{Q_H} compares the resulting $\rm Q_{HI}$ from the two feedback models considered.}

\section{Analysis}\label{results}

In this section we focus on the analysis of the simulations. In particular, we study the structural characteristics of the star cluster (such as mass function, virial status, mass segregation, escapers, binaries) in the three different runs, to understand the role of feedback (FB) in shaping the star cluster itself.

\begin{figure}
\includegraphics[width=\columnwidth]{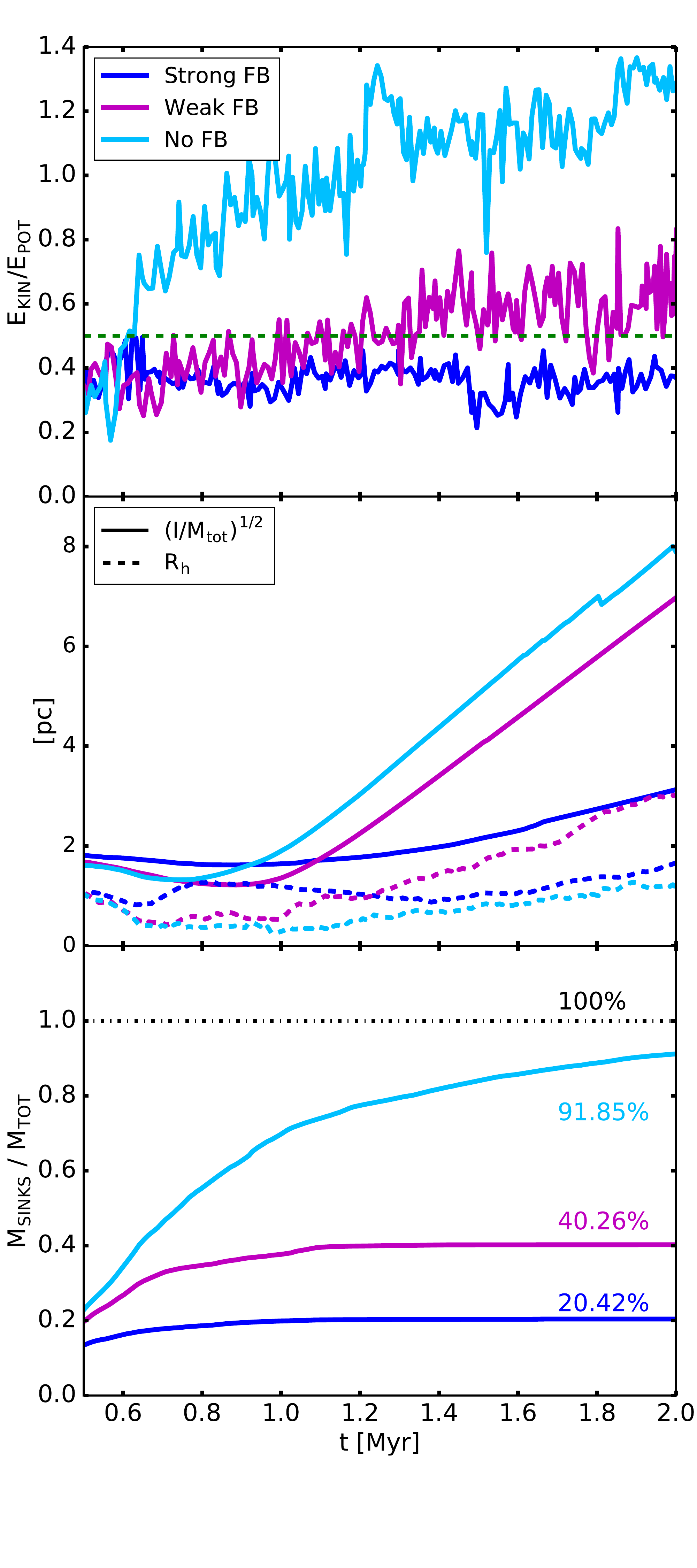}
\caption{Top panel: ratio of total kinetic to potential energy of the sinks (virial ratio, $\rm E_k/E_p$), the dashed green line indicates the virial equilibrium. {\review Middle panel: size of the star cluster, in terms of half-mass radius (dashed line) and moment of inertia-derived radius (solid line). $I$ and $M_{SINKS}$ indicate respectively the moment of inertia and total mass of sinks.} Bottom panel: star formation efficiency evolution with time computed as the mass fraction in sinks (M$_{\rm SINKS}$ indicates the total mass in sinks, M$_{\rm TOT}$, the total initial mass of the gas cloud). The ionising radiation suppresses the formation of stars by clearing gas out of the cloud, and it increases the virial stability of the emerging star cluster.}
\label{fig:virial}
\end{figure}

\begin{figure*}
\includegraphics{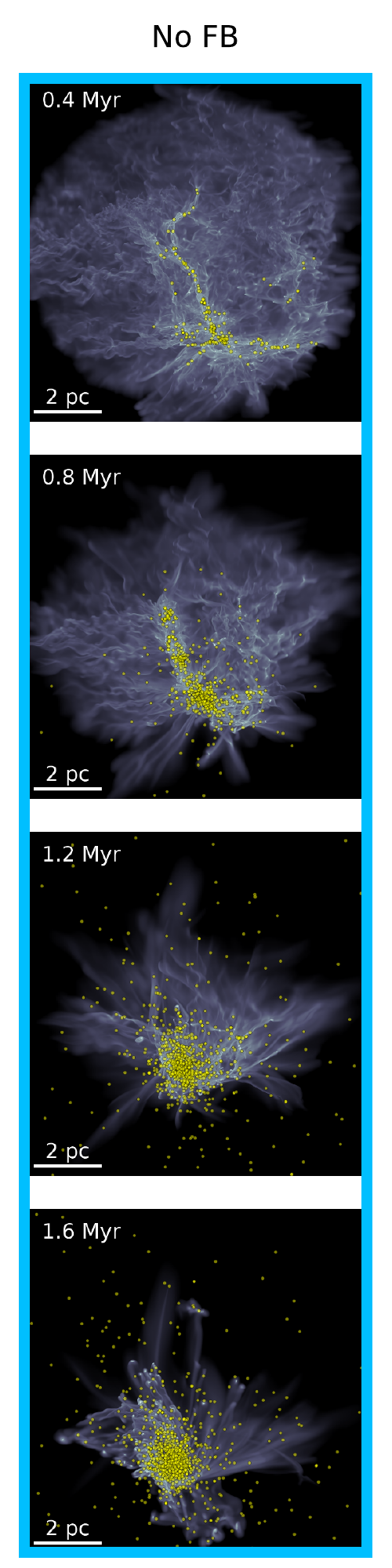}\;
\includegraphics{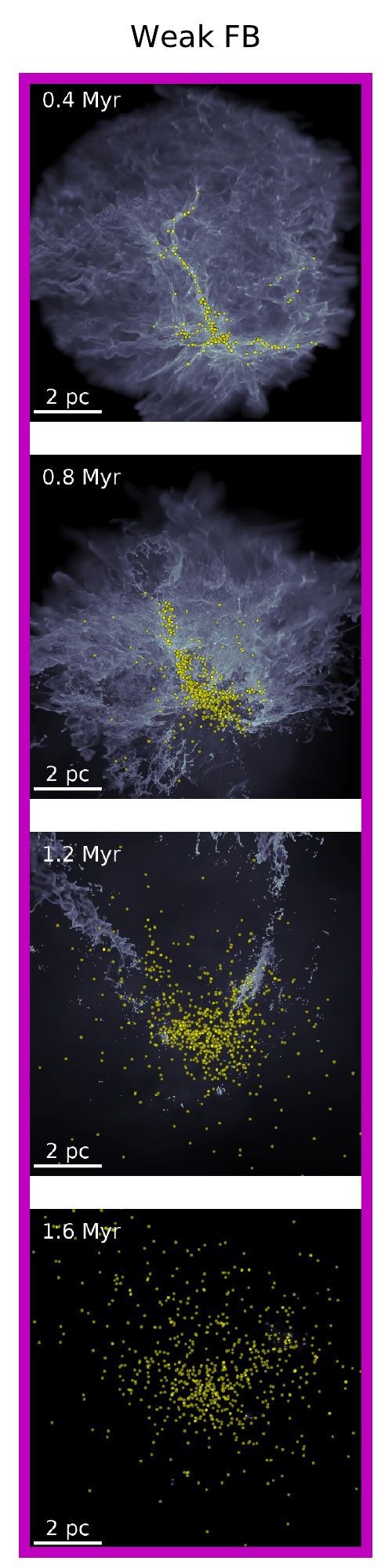}\;
\includegraphics{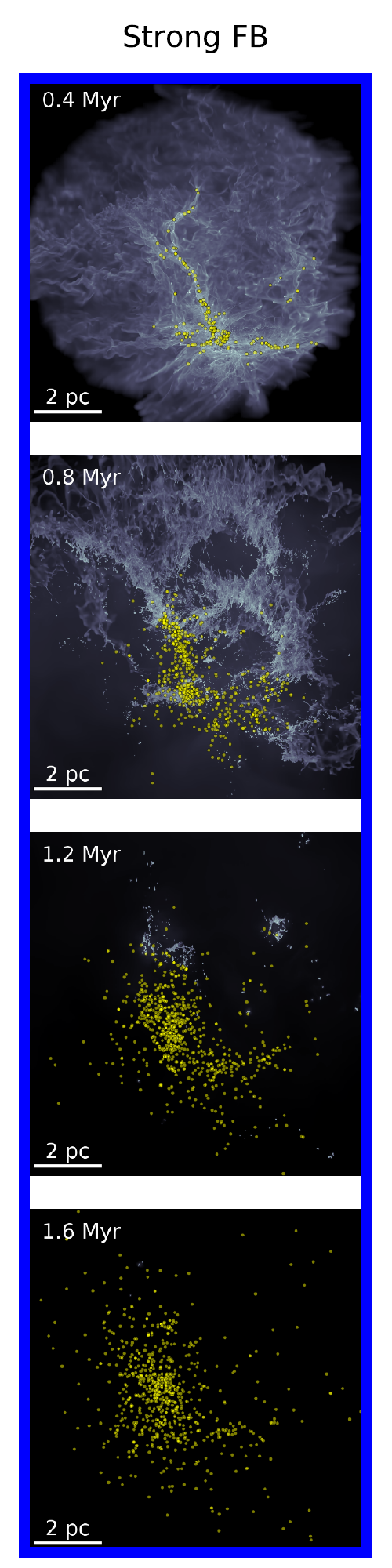}
\includegraphics[width=0.7\textwidth]{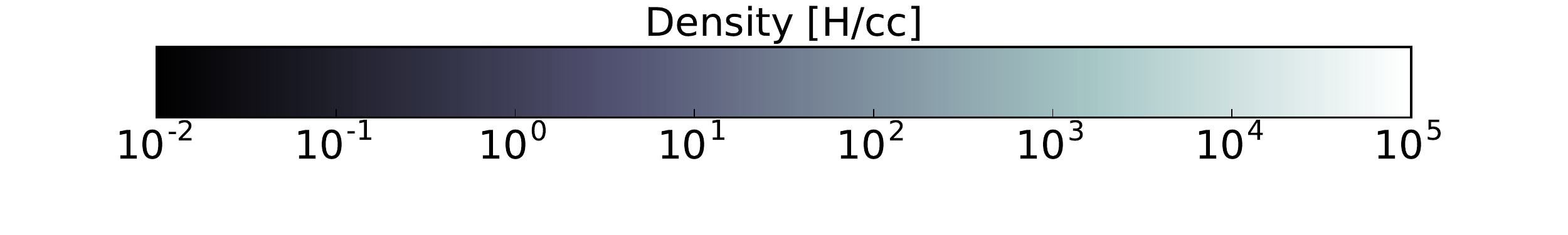}\;
\caption{Mass weighted line-of-sight projections of gas density for all three runs at different times. The strong feedback case in always denoted with dark blue colour, weak feedback case with magenta and the run without feedback with azur). Sink particles are indicated in yellow. }
\label{fig:dens}
\end{figure*}

\begin{figure*}
\includegraphics{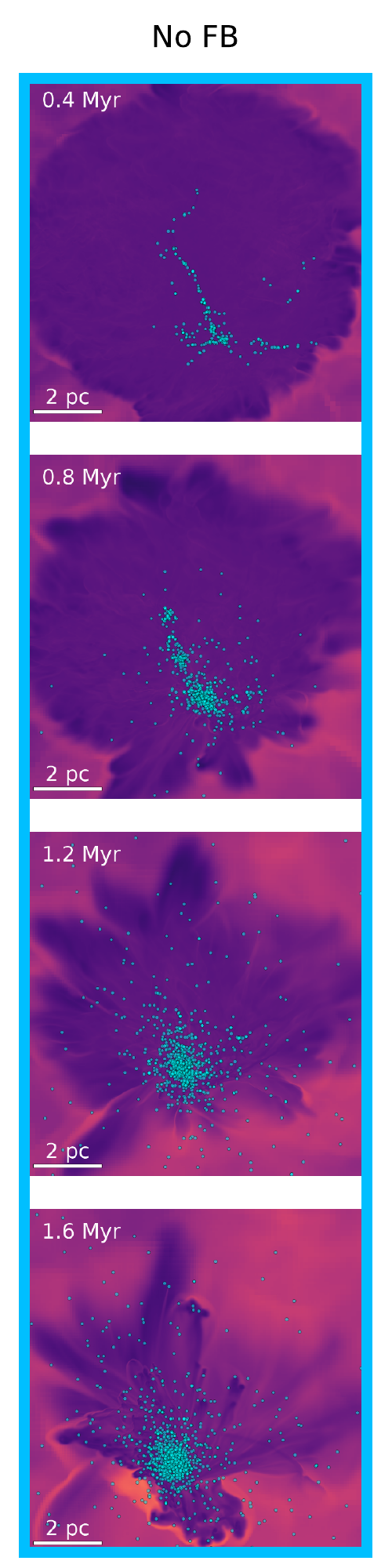}\;
\includegraphics{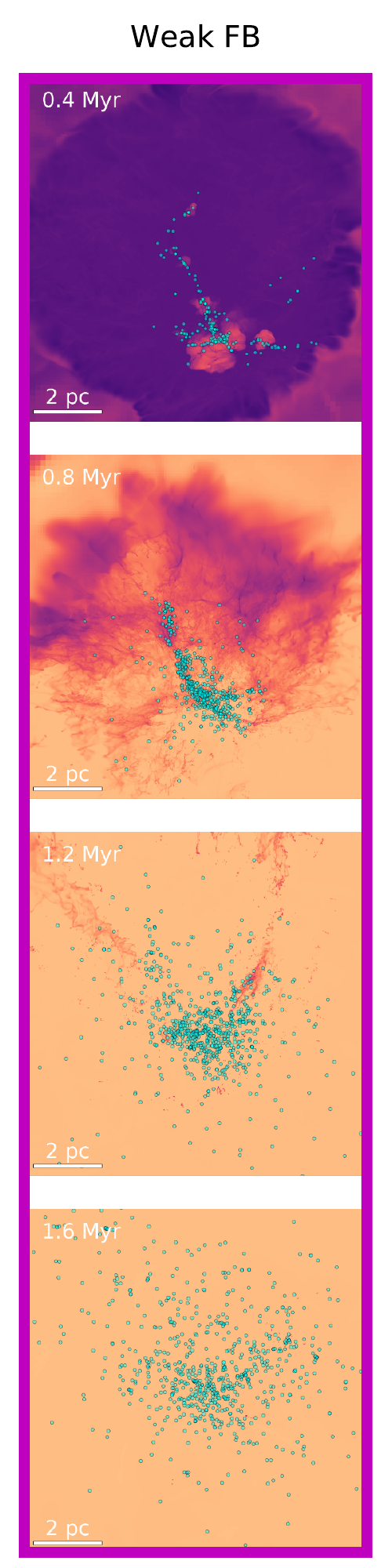}\;
\includegraphics{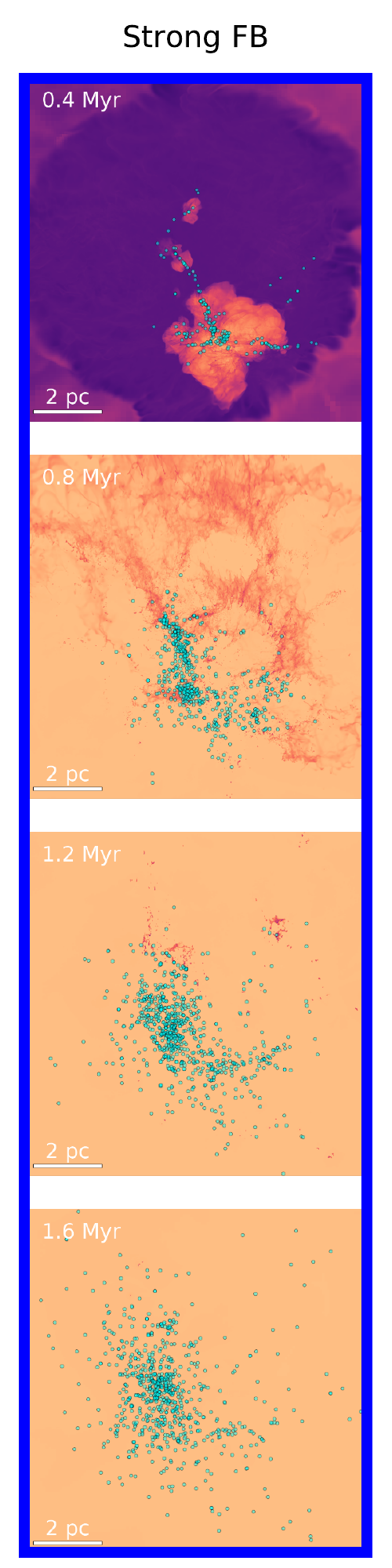}\;
\includegraphics[width=0.7\textwidth]{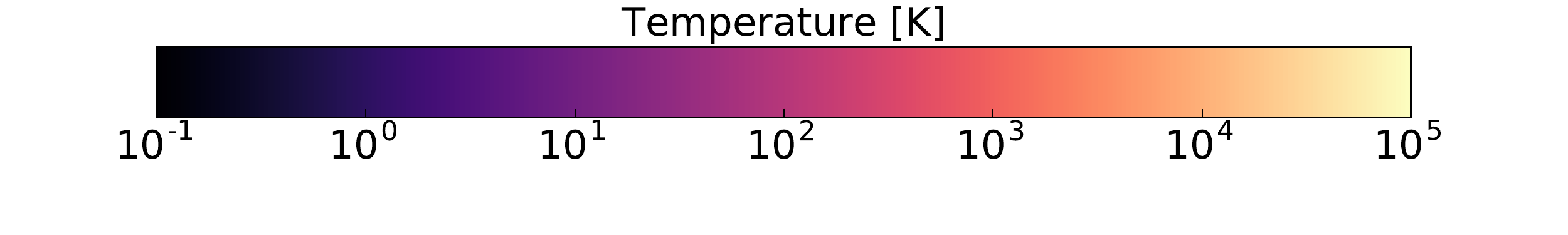}\;
\caption{Mass weighted line-of-sight projections of gas temperature for all three runs (strong, weak and no feedback) at different times. Sink particles are indicated in turquoise.}
\label{fig:temp}
\end{figure*}

Figure \ref{fig:virial} shows ratios of kinetic to potential energies of sinks (upper panel), {\review cluster sizes} and the SFE (lower panel) as a function of time. Focusing first on the SFE, the ionising radiation clearly has a major effect in suppressing star formation. In Figures \ref{fig:dens} and \ref{fig:temp} we demonstrate the effects of the radiation qualitatively, plotting time-sequences of gas density and temperature maps, to compare the strong, weak and no feedback cases.
The initial phase of the cloud collapse proceeds identically in the three cases. The cloud gravitationally contracts and starts forming filaments, where local overdensities allow the creation of stars, here represented by sinks (in yellow or turquoise, depending on the map). 

In the no-feedback case this contraction proceeds without resistance until, eventually all the gas is converted into stellar objects; from Figure  \ref{fig:dens} we can see how even in the latest snapshot the amount of dense gas is still high and by the end of the simulation time (2 Myr) the fraction of total mass still available in gas is $\rm \sim10\%$.
In general, we can notice how the final shape of the star cluster becomes more and more spherical with the simulation progressing. 
The gas temperature in the no-feedback case does not show huge changes throughout the collapse. 

In the weak-feedback case, stars emit ionising radiation and we now follow the photo-chemistry of Hydrogen. Differences with the no-feedback case start being visible around already 0.4 Myr in the temperature map, when the most massive stars in the lower part of the filament start emitting UV photons and cause the gas temperature to increase locally.  This bubble of hot gas becomes more and more extended since more stars are formed, accreting more gas. The neutral HI gets dissipated, due to the quick expansion of the HII region. At the end of the simulation, the star cluster is completely free of dense and neutral gas.  
The strong-feedback case is analogous to the weak-feedback case but the process of photoionisation and gas expulsion is much more rapid and violent, 
so as a result the star cluster is devoid of gas already at 1.2 Myr.   

\subsection{Virial properties}
In the top panel of Figure \ref{fig:virial} we show the evolution over time of the virial ratio of the star cluster, $\rm E_k/E_p$ (where $E_k$ and $E_p$ are respectively the total kinetic and potential energy of the sinks) for the three simulations. We do not consider snapshots before 0.5 Myr because {\review before this time there is still a large amount of gas mass which will become sinks, and therefore {\reviewtwo the stellar cluster} cannot yet be treated as isolated system.}  %the rate of sink formation is still to high and strongly influences the virial analysis. 

{\review As seen in the figure, the case without feedback is clearly super-virial, hence expanding. The two cases with feedback, instead, result in virial or even sub-virial state.}
This can be explained as a result of feedback, which halted the collapse of the cloud, ionising and dispersing the neutral gas. This determined the formation of a much less dense aggregation of stars than in the control simulation. In the run without feedback, the collapse proceeds unhindered and the new-born stars are immersed in a dense, highly-collisional environment, experiencing very strong close interactions. This inevitably leads to the ejections of many sinks and expansion of the cluster. {\reviewtwo The middle panel of Fig. \ref{fig:virial} clarifies the evolution of the size of the star cluster, considering both the half-mass radius (dashed line) and the global size obtained as $\sqrt{I/M}$, where $I$ is the moment of inertia of the cluster and $M$ is the total mass. From the plot it can be seen that the reference run is the more extended one, but at the same time half of its mass is very concentrated at the center. The expansion is then due to the escaping (massive) stars, not to a generally unbound cluster (a similar case was presented in \citealt{Kuznetsova2015})}. However, when feedback is included, its effect is to oppose this runaway collapse and allow the onset of a lower density regime, where the stellar distribution finds a stable configuration. 

It is interesting to notice how this result goes against traditional predictions (see the Introduction), which argue for a complete disintegration of the star cluster after a violent expulsion of gas. 
However, these often assume a fully formed star cluster still embedded in gas, which at some point gets ejected. In our case, stars are created while the gas is expelled in a self-regulating fashion. Therefore the virial status of the emerging star cluster changes along with the collapse. %In our simulations, the two main agents determining the internal dynamics and survival of the cluster are the virial ratio of the molecular cloud (highly subvirial) and the strength of the feedback: a very subvirial cloud produces a cluster too dense to survive, unless feedback slows down the collapse. 
{\review The outcome of our simulations results from the interplay between the highly subvirial initial virial ratio and the strength of the feedback adopted: a very subvirial cloud produces a cluster too dense to survive, unless feedback slows down the collapse.}
We also conclude that the star formation efficiency alone is not a good indicator of the survivability, as it is usually believed.

In the lower part of Figure~\ref{fig:virial} we show the fraction of gas transformed into stars. Stellar feedback is very efficient in stopping the collapse and lowering the SFE. In fact in the case with the strongest feedback the SFE halts at $\sim$20\% (while virtually unity for the control simulation). For a weaker feedback, we get a higher fraction. Despite the fact that in the simulation without feedback all gas is eventually transformed into stars, we stress that the outcome of the simulation is the dispersal of the emerging star cluster, while for the strong feedback case, which results in a very low star formation efficiency, the outcome is a stable (or even subvirial)  star cluster.

\subsection{Mass function}

\begin{figure*}
\includegraphics[width=\textwidth]{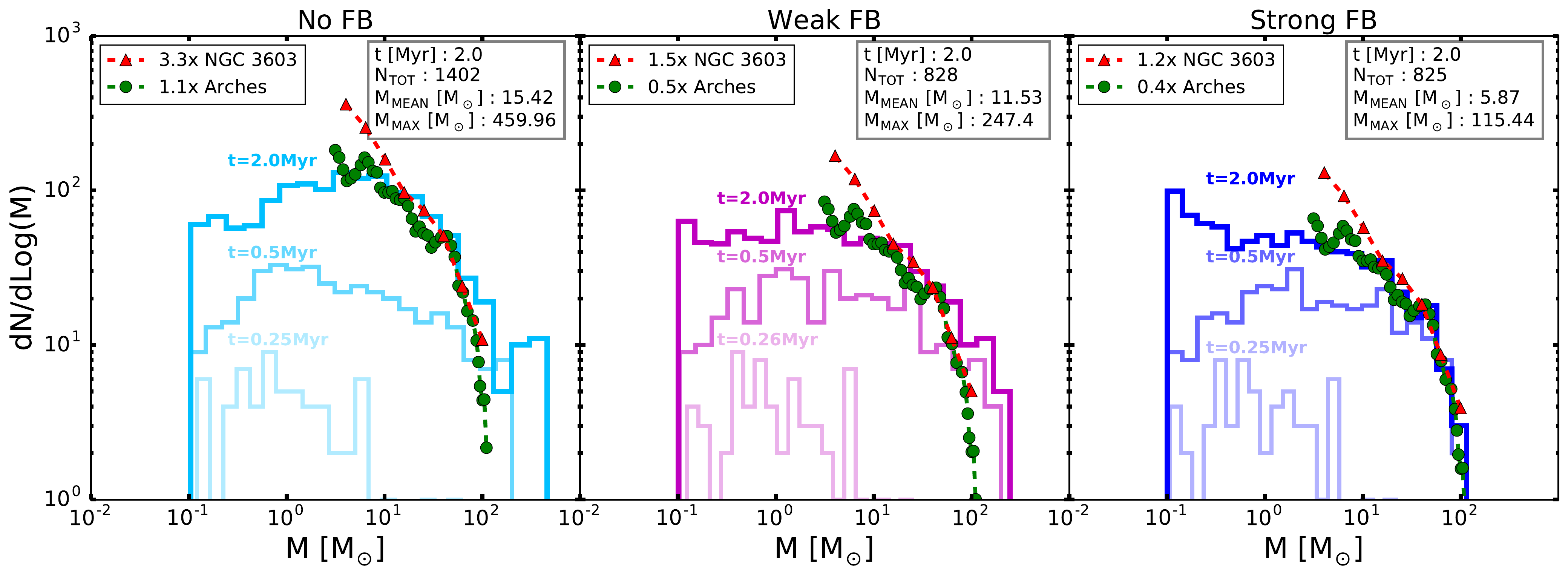}\;
\caption{Mass function profile for all 3 feedback regimes. The thicker and darker lines indicate for every run the mass function profile at t=2 Myr. In the top right box of every subplot the we indicated the total number of sinks, the average and maximum sink masses at t=2 Myr. In every subplot we plot also the mass functions at t=0.5 Myr and t=0.25 Myr in paler colours. The green dots correspond to the normalised observational data for the Arches cluster \citep{Stolte05}, the red triangles the same for NGC 3603 \citep{Pang13}. The normalisation factors are reported in the top left box. Our strong feedback case compares best with the normalised observations, although all models fail to reproduce the steep curve towards lower masses in NGC 3603.}
\label{fig:mf_t}
\end{figure*}

In Figure~\ref{fig:mf_t}, we plot the stellar mass function for all the feedback cases we have considered and at different times.
In the run without feedback, our mass function peaks at a relatively large mass of $\sim$10 \msun and shows a strong accumulation of very massive stars at the high mass end, 
with the mean sink mass being around 15 \msun{}  and the most massive sink reaching 460 \msun. This is due partly to our limited resolution (see later) and to the lack of feedback to limit the maximum stellar mass.
In the weak feedback scenario, the maximum mass is lower, around 250 \msun{} and the mass function flattens, with a slight increase of very low mass stars (close to our resolution limit of 0.1 \msun).
The trend gets even clearer if we look at the case with strong feedback, where there is a significant peak of stars with mass around 0.1 \msun{} (corresponding to the sink seed mass) 
and the most massive star is now around 120 \msun. We observe in the simulation that this excess of low mass stars close to the resolution limit is caused by the fragmentation of the outer dense shells
of HII regions.

Looking at the mass function at earlier times (specifically, t=0.25 Myr and t=0.5 Myr, paler lines in figure 3), it is clear that the onset of the sink mass function proceeds similarly in the three cases. 
It is mainly the final mass distribution that shows visible differences between the feedback and no-feedback cases. 
To summarise, these are 1) the high-mass cut-off due to feedback effects that stop accretion onto massive sinks, 
2) a peak at the low-mass end, due to fragmentation of dense gas around HII regions.

%gas dispersal 

%old version, when sinks < 0.1 Msol where there
%The main difference between the run with and without feedback lie in the amount of gas reservoir, sinks can accrete from. By looking at the mass function at earlier time (specifically, t=0.25 Myr and t=0.5 Myr, paler lines in figure 3) we can try to understand better how the "growth" of the mass function proceeded. The very first stars formed are intermediate-mass stars with mass close to the Jeans mass expected according to the threshold sink density used in the simulation (t=0.25 Mry). Later on the sinks accrete the surrounding gas and we can see how the mass function is unbalanced towards higher masses (t=0.5 Myr). Then, in the cases where feedback is turned on, the accretion process eventually decreases and the gas get ionised: therefore the new stars born in gas-depleted areas, where there is not much material to accrete, hence they keep being very low-mass stars (M<0.1\msun).

\subsubsection{Comparison to observations}

It is very instructive to compare the results of our simulations to available observations. 
We choose to consider NGC~3603~YC and the Arches cluster, since they are among the youngest (< 2~Myr) well-studied star clusters, {\review part of large, still gas-rich, regions}.

NGC~3603~YC (also known as HD 97950) is a very compact and massive young star cluster at the centre of the vast homonym HII region. 
It is composed of three Wolf-Rayet stars and around 40 O-type stars, a dozen of which resides in the very central part of the core, within less than 1~ly from the centre \citep{Drissen95}. 
\cite{Harayama08} estimated the total mass to be between 1 and $2 \times$ 10$^4$ \msun. The H-R diagram in \cite{Melena08} reveals the presence of at least 15 stars with masses greater than 60 \msun.
The most massive stars in the cluster seem to be coeval with ages between 1 and 2~Myr \citep{Kudryavtseva12,Melena08,Stolte04}. 
However, the age spread between the pre-main-sequence stars \citep{Beccari10} and the slightly older stars in the cluster outskirts \citep{Sung04} suggests a possible extended star formation scenario. 

The Arches cluster is considered to be the densest cluster in our Galaxy. 
It also falls in the category of so-called starburst star clusters. It is located near the Galactic centre and its age is estimated to be around 2~Myr.
Its total mass is estimated to be around 2 $\times$ 10$^4$ \msun \citep{Espinoza09} and it contains 160 O-stars and 13 Wolf-Rayet\footnote{This is about 5\% of all known Wolf-Rayet stars in the Milky Way \citep{Figer02}.} \citep{Martins08,Figer04}.

For NGC~3603~YC, we considered the mass function results published by \cite{Pang13} and for the Arches the one published by \cite{Stolte05}. 
To derive the mass function of NGC~3603~YC, the authors considered stars in absolute V-magnitude bins and then derived the correspondent masses using the isochrone models from \cite{Lejeune01} 
for high mass stars and \cite{Siess00} for low mass stars.
Their mass bins have a logarithmic size of 0.2.
The data were corrected both for incompleteness and foreground stars contamination and include all stars within 60'' ($\sim$ 2 pc).

\cite{Stolte05} derived the present day mass function of the Arches cluster by converting the K-band magnitudes from the corrected color-magnitude diagram into masses using a 2 Myr Geneva main-sequence solar metallicity isochrone from \cite{Lejeune01}.
They also binned their data using logarithmic intervals of size 0.2 and they computed the mass function 10 times, each time shifting the bins by 0.02. 
The final present-day mass function was created by averaging all the points from these 10 mass functions and takes into account all stars within 0.4 pc.
%2 and 0.4 pc are considered the clusters dimensions

Comparing these observational data to our simulations is not trivial, since we do not know the SFE of the parent clouds of both NGC~3603~YC and the Arches. 
The targeted clusters have about twice the mass of our simulated ones from the feedback runs, but roughly equal to the one in our no-feedback simulation. 
If the true SFEs of the observed star clusters were very low, say 10\%, this would imply that the original clouds would be as massive as $10^5 M_\odot$, 
which is computationally too expensive to simulate at the current resolution and with our radiation solver. 
Therefore, we decided to re-normalise the observations. The normalisation factors are computed requiring that the mass bin at $15~M_\odot$ in the two observational datasets
have the same value, equal to that of our simulated data set. The normalisation coefficients for the Arches dataset are 0.4, 0.5 and 1.1 with respect to the strong, weak and no feedback cases, respectively,
while the normalisation coefficients for the NGC~3603 dataset are 1.2, 1.5 and 3.3 with respect to the strong, weak and no feedback cases, respectively

In Figure \ref{fig:mf_t} we compare these renormalised observed mass functions to our simulated ones. 
Renormalised observational data are showed with red triangles (NGC~3603) and green circles (Arches). 
The best agreement, especially at the high-mass end, is obtained with the strong feedback (after renormalisation). 
The weak and no-feedback runs clearly produce too many very high-mass stars. The agreement is worse at lower masses, especially below $10~M_\odot$.
As we explain below, we believe this is due to our limited resolution. 

\subsubsection{Slope of the mass function}

\begin{figure*}
\includegraphics[width=\textwidth]{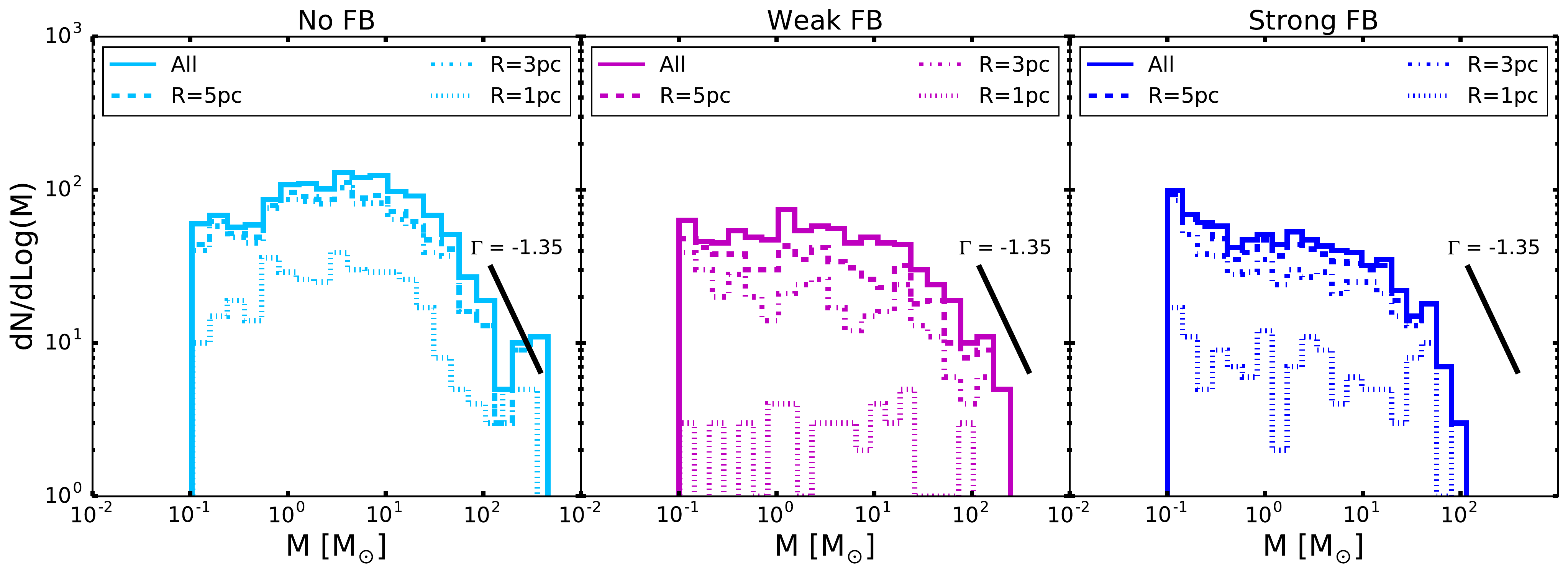}\;
\caption{Mass functions at different radii, R=1, 3, and 5 pc. The solid thick line refers to the total mass function (all sinks included). In black the Salpeter slope ($\Gamma$ = -1.35) is indicated as a reference. In all cases the slope of the simulated mass functions is flatter than the Salpeter one. Moreover, the slope steepens
with radius, especially in the feedback cases, indicating a higher concentration of massive stars in the centre.}
\label{fig:mf_r}
\end{figure*}

The previous analysis was carried out considering all the sinks in the simulation box. We now study the mass function dependency with radius. 
In Fig. \ref{fig:mf_r}, we show the mass function taking into account only sinks within specific radii\footnote{Unless otherwise stated, the radius is always considered respect to the centre of density of the system defined as in \cite{Starlab}.}%but with 32 particles 
, namely 1, 3 and 5~pc, and for all three feedback regimes. 
The last radial bin contains 92\%, 74\% and 88\% of the simulated sinks respectively for strong, weak, no feedback. 
The solid curve corresponds to the whole box, or a radius of 10~pc.
Although the mass function appears to be independent of radius for the no feedback case, it looks clearly flatter in the inner parts and steeper in the outer parts for the two feedback cases.
\cite{Pang13} showed that a similar effect is present in NGC~3603: the slope of the mass function steepens with radius, indicating that the most massive stars are mostly concentrated in the centre. 
This feature is generally explained by mass segregation. We will develop this topic in the next section.

If we now quantify the slope of the mass function, we found that all our simulations show a slope ($\Gamma$) much flatter than that of the Salpeter IMF (i.e. $\Gamma$=-1.35), 
depending sensitively on the range of masses used to compute it (see Fig. ~\ref{fig:mf_r}){\review, which means the mass function is probably not a power-law all in all.
A shallower slope than the Salpeter} is also the case for observed young and embedded star clusters. 
NGC~3603, for example, has $\Gamma = -0.88 \pm 0.15$, considering only log(M/\msun)~>~0.6 for completeness reason.
For the Arches, \cite{Stolte05} detected a change in the slope of the mass function at about 6~$M_\odot$, hence they fitted the mass function in the range  log(M/\msun)~>~0.8. 
The resulting value was measured to be $\Gamma=-0.86 \pm 0.15$. Both these clusters have slopes flatter than the Salpeter slope, 
which seems to be in general a distinguishing feature of young starburst clusters.

The origin of this discrepancy from the Salpeter slope is probably due to many reasons. 
On the simulation side, \cite{BertelliMotta2016} showed that the simulated IMF can be affected by resolution, with the peak or turn-over mass depending directly on it. 
The higher the resolution, the lower the turn-over mass, which implies a progressive steepening of the mass function with increasing resolution. 
These authors estimated that the peak mass is roughly $\sim 30$ times the minimum Jeans mass, 
which is our case corresponds to about 4.5 $M_\odot$, and agrees quite well with our no-feedback case. 
{\review Studying the formation of low-mass protostars in radiative feedback simulations, \cite{Bate2014a} obtained IMF profiles with slopes compatible with the Salpeter prescription.}

So, resolution effects are likely a cause of the low value for $\Gamma$ in our simulations in the intermediate mass range log(M/\msun)~<~1.
Moreover, feedback inevitably plays a role in all this, lowering the number of stars in the intermediate-high mass range, therefore contributing to an even shallower slope. 
At larger masses, on the other hand, recent theories of turbulent cloud collapse argue for an asymptotic Salpeter slope \citep{Hennebelle2008, Hopkins2012h}.
This could be consistent with our simulated star clusters, but also with the observed ones, without being very conclusive, reminding us that the story is probably not so simple. 
 
%It is quite possible that the mass function of these observed objects is not yet the final one, due to their very young age. 
%The final IMF could evolve into a steeper function by dynamical and stellar evolution.
%Another theory for the IMF is based on the self-similar fragmentation of molecular clouds into clumps and clumps into stars, leading to a power law mass distribution with $\Gamma=1$  \citep{Oey12}.
%Yet another explanation has been designed using competitive Bondi accretion on random seeds, leading also to a $\Gamma=1$ mass function \citep{Zinnecker82}. In these two theories,
%the Salpeter slope was obtained by identifying massive stars as rare events that cannot form in the low-mass available gas clumps of the cluster \citep{Oey12}.

\subsection{Mass segregation}\label{massseg}

We have already introduced mass segregation in the previous section to explain a steepening of the slope of the mass function as a function of radius.
We now analyse our simulations with more traditional  tools to quantify mass segregation in star clusters.
A star cluster is considered to be mass segregated when the massive stars are more centrally concentrated than the lower mass stars. 
The main question related with mass segregation is whether it has a primordial or a dynamical origin. 
Mass segregation can indeed be the result of two or three body interactions between stars (dynamical) 
or the direct outcome of the star formation process within the gas cloud itself (primordial).
Our simulations are ideal experiments to try and answer this question.

The problem of comparing the mass function for different radii to characterise mass segregation is that we need to define unambiguously the centre of the star cluster, which is a difficult task. 
\cite{Allison09} introduced the Minimum Spanning Tree (MST) to quantify the degree of mass segregation in a star cluster.
The MST is defined as the shortest path connecting all points, which does not contain any closed loop. 
We used the routine included in the {\ttfamily csgraph} module of {\ttfamily scipy}, which implements the MST according to Kruskal's algorithm \citep{Kruskal56}.   

We followed \cite{Allison09} prescription to quantify mass segregation using the MST. 
We computed the length, L$_{\rm massive}$, of the MST of the N most massive stars and compared this to the average length of the MST of N random stars in the cluster, or L$_{\rm random}$. 
L$_{\rm random}$ was calculated by picking 1000 random sets of N stars, in order to have a small error on the dispersion $\rm \sigma$.
Mass segregation is quantified using the {\it Minimum Spanning Tree Ratio} $\rm \Lambda_{MSTR}$ defined by \cite{Allison09} as $$\rm \Lambda_{MSTR} = \frac{L_{random}}{L_{massive}} \pm \frac{\sigma}{L_{massive}}.$$ 
For $\rm \Lambda_{MSTR} \sim 1$, the distribution of massive stars is comparable to that of all stars. 
For $\rm \Lambda_{MSTR} > 1$, massive stars are more concentrated, a clear sign of mass segregation. 
The larger $\rm \Lambda_{MSTR}$, the more pronounced is the mass segregation. 
 
This method was already adopted by \cite{Parker14, Parker15} to analyse the dynamical evolution of star forming regions, starting from the final states of the SPH simulations by \cite{Dale12a, Dale12b}.
%In \cite{Parker15} they compared different methodologies to compute mass segregation and conclude those techniques give different answers to the debate upon primordial/dynamical origin. 
Using $\rm \Lambda_{MSTR}$ for their $N=10$ most massive stars, they found in their no-feedback simulation a strong primordial mass segregation with $\rm \Lambda_{MSTR} \simeq 5$, 
which disappears after 3~Myr due to stellar evolution and reappears at the same level after 8~Myr due to dynamical interactions between the cluster members. 
However, in their feedback simulations that include winds and photoionisation, they did not detect any mass segregation, with $\rm \Lambda_{MSTR} \simeq 1$ at all times.

\begin{figure}
\includegraphics[width=\columnwidth]{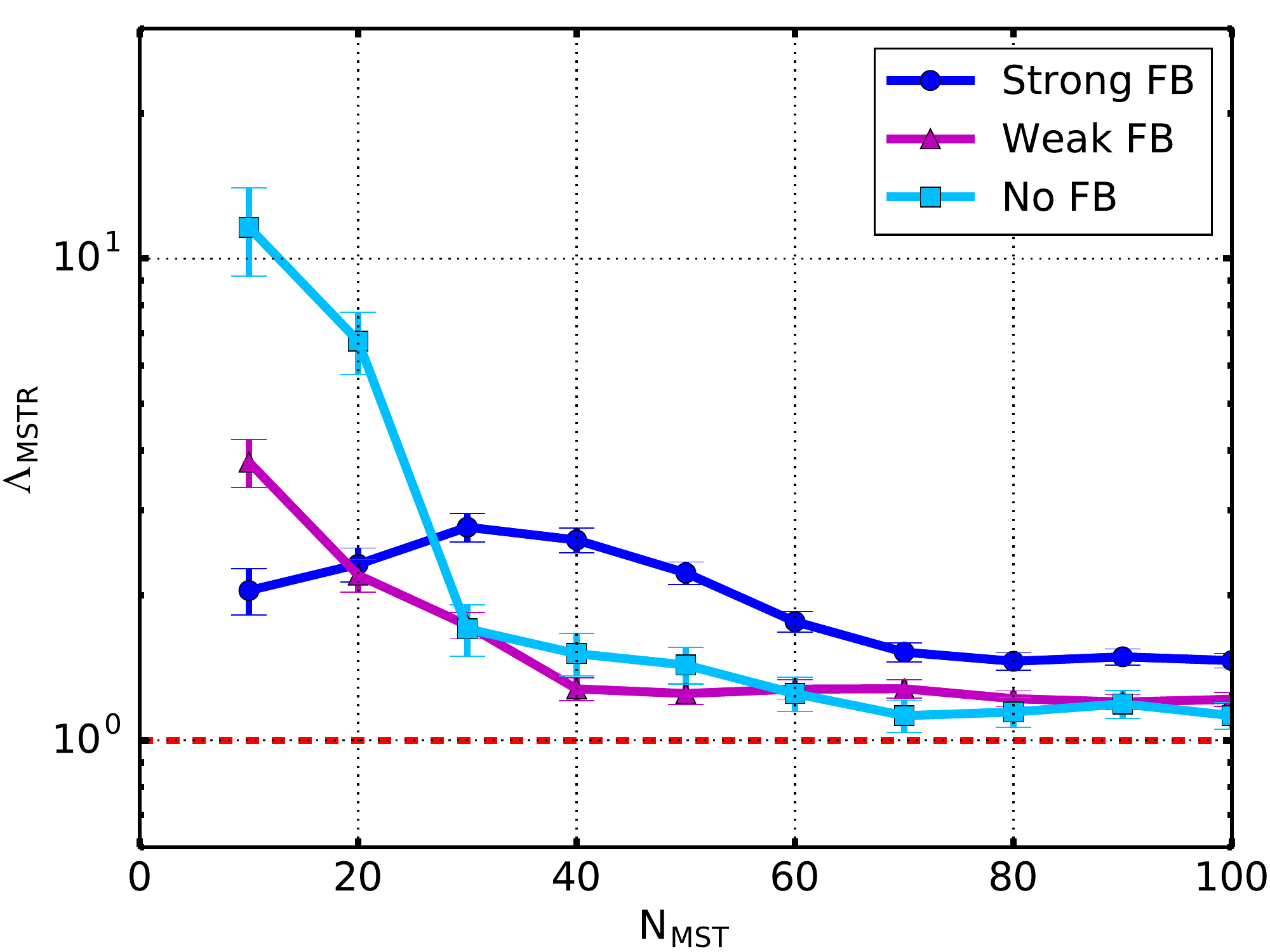}
\caption{Minimum spanning tree ratio $\rm \Lambda_{MSTR}$ against number of stars used to calculate the length of the tree. The red dashed horizontal line indicate the value $\rm \Lambda_{MSTR}$=1, meaning no mass segregation. The vertical bars correspond to 1 $\rm \sigma$ error of $\rm \Lambda_{MSTR}$. All three cases show some degree of mass segregation.}
\label{fig:mst_n}
\end{figure}

In Figure~\ref{fig:mst_n}, we plot $\rm \Lambda_{MSTR}$ as a function of N$_{MST}$, the number of stars we use for the spanning tree, at t~=~2Myr. 
We  include in our analysis all stars up to an outer radius of 7.7 pc, 9.3 pc and 9.8 pc, corresponding to the distance from the centre of the cluster 
of the most external bound star, in the strong, weak, and no feedback cases respectively.
This is done to prevent extreme outlier stars to dominate the calculation of the random spanning tree.
Our data point with $N=10$ corresponds to the estimator used in \cite{Parker15}.

All three cases show some degree of mass segregation. 
Our no-feedback case is strongly mass segregated for N=10 with $\rm \Lambda_{MSTR} \simeq 10$, and is still significantly segregated for N=20 with $\rm \Lambda_{MSTR} \simeq 5$.
The signal however disappears for $N\ge30$. The strong feedback case shows the weakest mass segregation for $N=10$ with $\rm \Lambda_{MSTR} \simeq 2$, but the segregation signal
is still detectable up to $N=60$. The weak feedback case lies in between the two other cases. 

The two crucial pieces of information Figure~\ref{fig:mst_n} provides are 1) the degree of mass segregation of the cluster, namely the value of $\rm \Lambda_{MSTR}$ and 
2) the extent of mass segregation, namely the maximum number of stars that are mass segregated.
From our results, two different situations emerge. In the no feedback case (and to some extent in the weak feedback case), 
only a handful of super-massive  stars are tightly concentrated at the centre. Only those most massive stars are mass segregated. 
The high stellar density is supported by the high measured values of $\rm \Lambda_{MSTR}$. 
This population of massive stars forms effectively a sub-cluster at the centre of the main cluster, that keeps contracting and decouples dynamically from the rest, 
transferring its kinetic energy to less massive stars that are ejected (see next Section). 

%Sometimes, equipartition, therefore complete extended mass segregation, can not be reach (cfr. Spitzer's instability). 
%The weak-FB case is probably a milder example, indicating that feedback has a counter effect on the onset of the instability.

On the contrary, in the strong feedback case, photo-ionisation feedback is efficient enough to halt the collapse of the gas, limiting the number density of massive stars. 
This prevents the formation of an independent self-gravitating system within the cluster itself. 
This translates into a lower degree of mass segregation and at the same time a higher number of stars being mass segregated.

\begin{figure}
\includegraphics[width=\columnwidth]{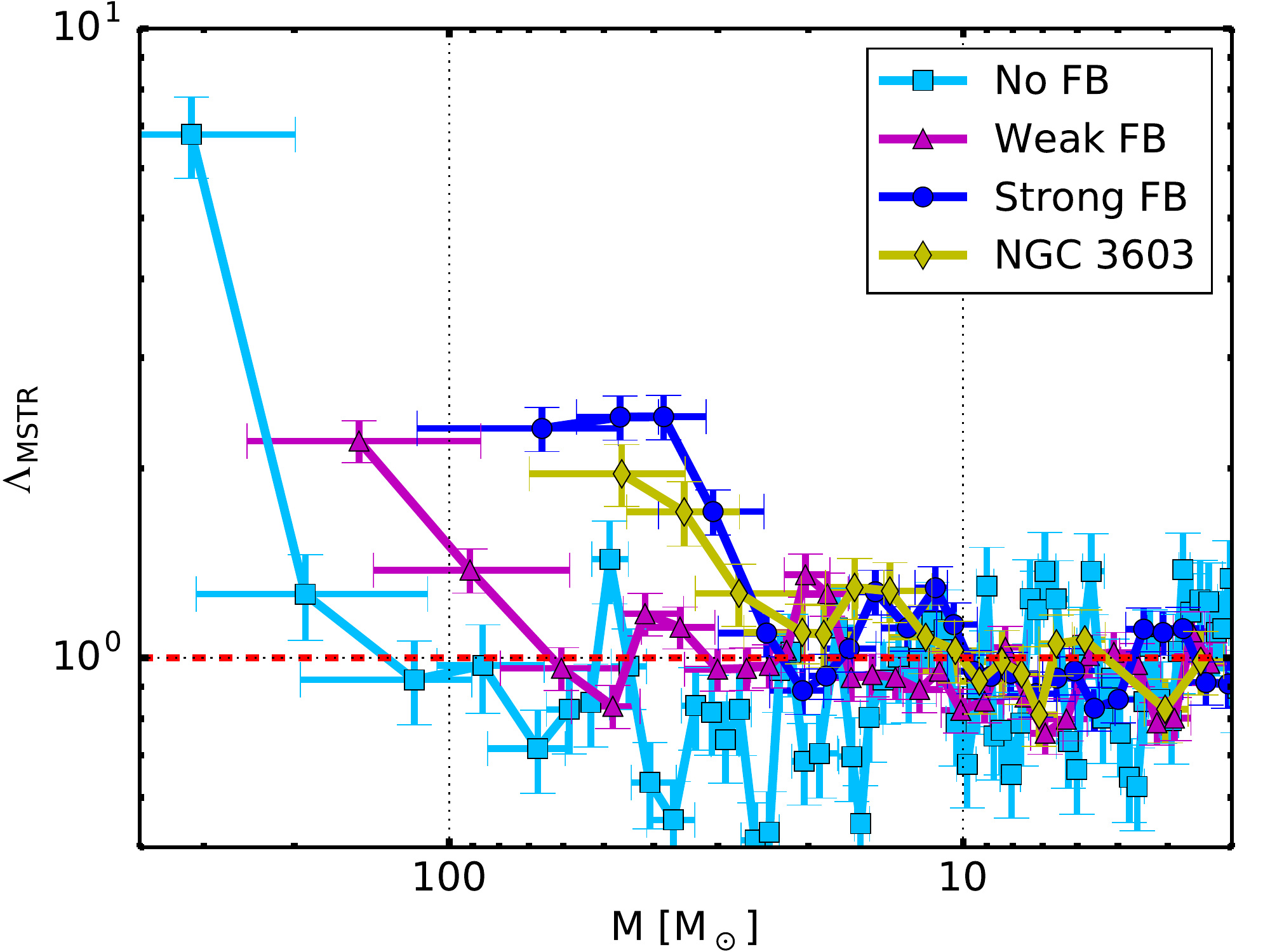}
\caption{Minimum spanning tree ratio $\rm \Lambda_{MSTR}$ versus stellar mass. The red dashed horizontal line indicate the value $\rm \Lambda_{MSTR}$=1, meaning no mass segregation. The vertical bar corresponds to 1 $\rm \sigma$ error of $\rm \Lambda_{MSTR}$. Horizontal bars show the mass interval covered by every group of 20 stars. Note that the horizontal line associated to the first data point for the non-feedback case extends to the left until $\sim$ 400 \msun. Observations are indicated in yellow. All curves show a similar behaviour, even if shifted towards higher masses. The best agreement with observations is provided by the strong feedback simulation.}
\label{fig:mst_m}
\end{figure}

In order to compare with observations, we plot $\rm \Lambda_{MSTR}$ as a function of the stellar mass (Fig.~\ref{fig:mst_m}). 
Following \cite{Pang13}, we sort the stars by their mass and then consider blocks of 20 stars moving in steps of 10 stars, such that the data partially overlap.
For example,  the first 20 stars in the weak feedback case (magenta line in Figure \ref{fig:mst_m}) cover the range 200 to 80\msun{} in mass, 
the second mass group goes from 130 to 60\msun{}, etc. The mass interval considered is indicated by horizontal bars in the plot. For every bar a marker denotes the mean mass of the interval.   

The three profiles of $\rm \Lambda_{MSTR}$ versus mass in Figure~\ref{fig:mst_m} look qualitatively similar, but they are shifted to higher and higher masses with increasing feedback strength. 
The no feedback case shows mass segregation only in the first bin (M>200\msun) with an amplitude much larger than unity. %the other spikes in the no fb case are effect of very close binaries
For the weak feedback case, only stars down to a mass of 60~\msun{} are weakly segregated, with an amplitude of 2, and for the strong feedback case, the transition goes down to 30~\msun. 

\begin{figure}
\includegraphics[width=\columnwidth]{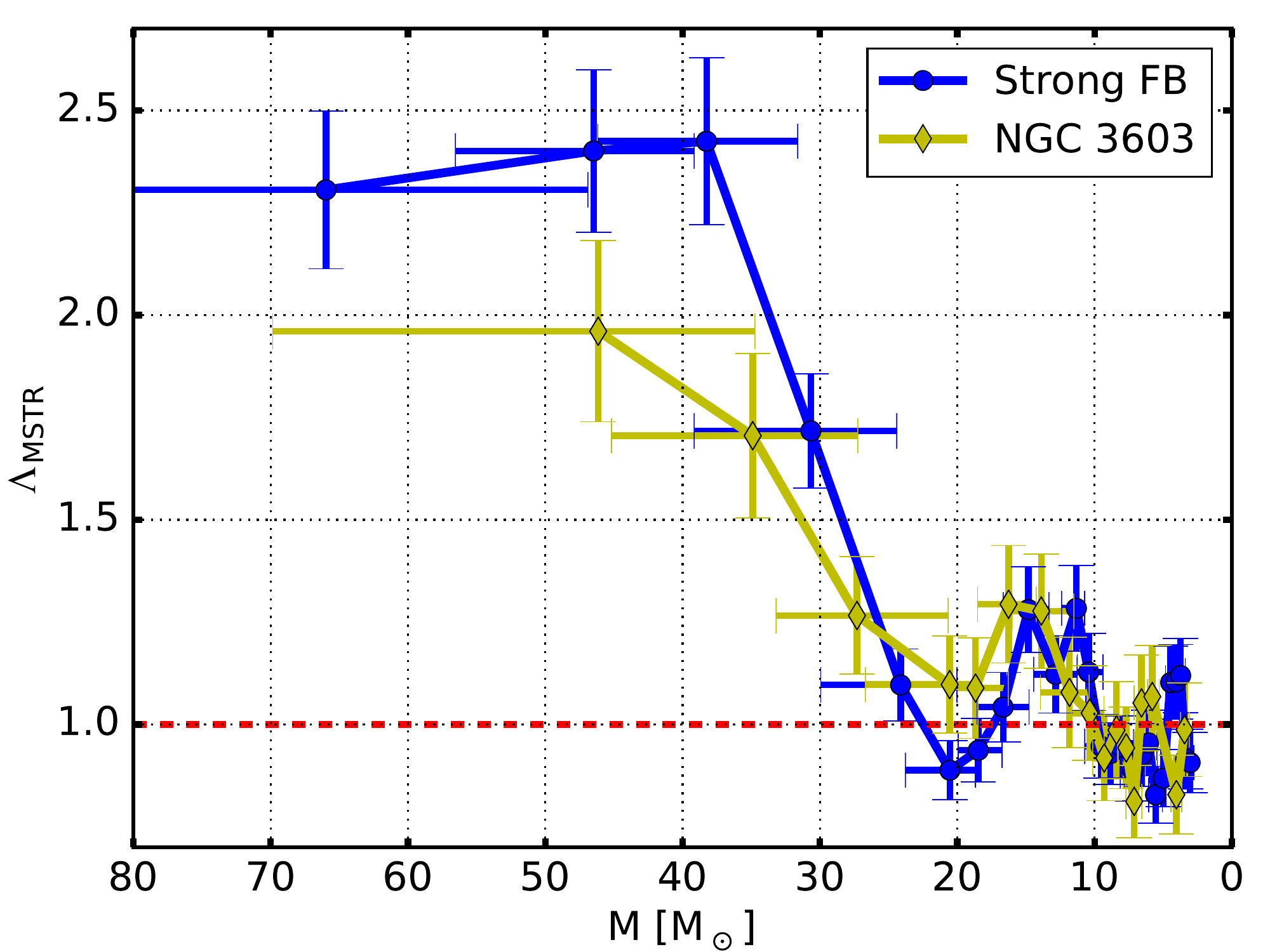}
\caption{Zoom-in plot of Figure \ref{fig:mst_m}. The comparison here is only between strong feedback case and NGC 3603 YC.}
\label{fig:mst_m_strong}
\end{figure}

In Figure~\ref{fig:mst_m}, we compare our simulations to the data of \cite{Pang13} on NGC~3603 (yellow points). A very good agreement is obtained with the strong feedback case. 
In Figure \ref{fig:mst_m_strong}, we plot only the strong feedback case and the observations using a linear scale in mass to allow a better comparison and to outline the very good quantitative match
between our model and the observed segregation, both in terms of amplitude and of transition mass. 

Despite being young, NGC~3603 shows already a clear signal of mass segregation. 
This is not an isolated case. There is also strong evidence of mass segregation in the Orion Nebula clusters, but also in the Arches, NGC~6611, NGC~2244 and NGC~6530, 
to name a few \citep{Hillenbrand1998, Stolte2002, Bonatto2006a, Schilbach2006, Chen2007}.
The origin of the mass segregation in these clusters is still an open question (primordial or dynamical). 

\cite{Pang13} proposes for NGC~3603 a dynamical origin. 
Using analytical arguments, they show that the cluster dense core could dynamically segregate in one crossing time down to a mass of 30~\msun.
To test this hypothesis, we have performed our clustering analysis at earlier times and find no indication of mass segregation for massive stars. 
We have estimated the local two-body relaxation timescale of the densest part of the cluster ($r < 2$~pc) and find it to be less than 0.5~Myr for all 3 cases,
supporting our claim that dynamical friction can cause mass segregation after 1~Myr.

To quantify further the structure and morphology of our star clusters , we have used another statistical indicator called the Q parameter \citep{Qparam}.
Q is defined as the ratio between the normalised mean edge length $\rm {\overline{m}}$ of the MST of all stars in the cluster 
and the normalised correlation length $\rm {\overline{s}}$ of the same stars\footnote{The correlation length is defined as the mean separation between stars in the cluster.}. These parameters taken separately cannot distinguish between a smooth, radially concentrated distribution and an extended, fractal distribution, but their ratio can \citep{Qparam}.  A cluster with Q > 0.8 is smooth and centrally concentrated, while if Q < 0.8, it is extended with a fractal distribution. 

\begin{figure}
\includegraphics[width=\columnwidth]{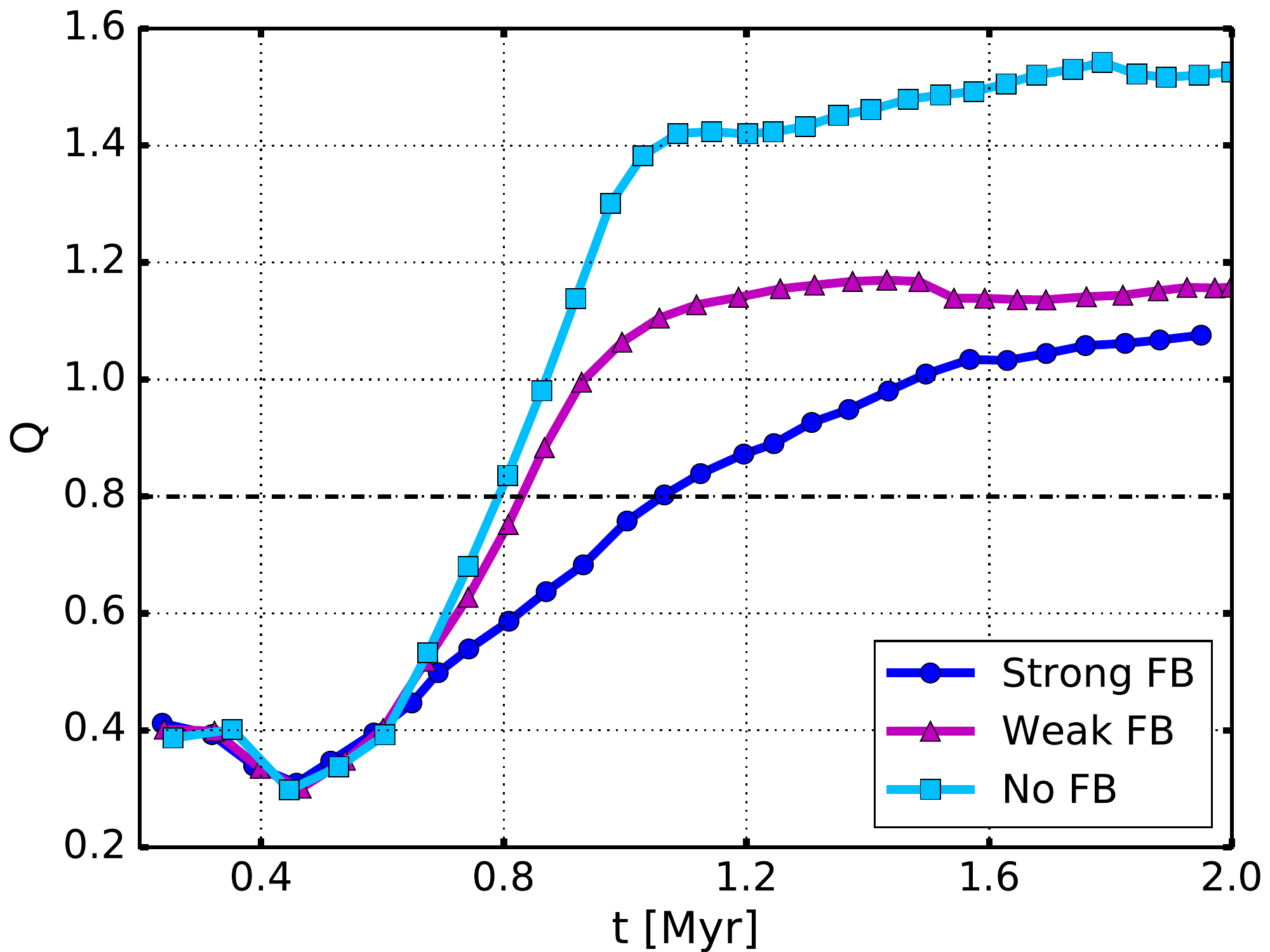}\;
\caption{Evolution with time of the Q parameter. The dashed horizontal line correspond to Q=0.8. This value discriminates between centrally concentrated (> 0.8) and fractal (< 0.8) spatial configuration. The simulations with feedback preserve substructures longer than the control run.}
\label{fig:q}
\end{figure}

In Figure~\ref{fig:q}, we show the evolution of the Q parameter with time. In all our simulations, the star cluster is initially fragmented and extended. 
The no feedback case rapidly evolves towards a more spherical and centrally concentrated distribution with $Q \simeq 1.5$, 
while in the two other cases, the transition is slightly slower, supporting a longer survival of substructures, and reaches a smaller maximum value with $Q \simeq $~1.1 and 1.2.
This supports a scenario in which gravitational collapse together with stellar dynamical interactions progressively erase the initial conditions in the gas cloud 
and build up a dense and spherical star cluster. In this context, feedback acts as a delay mechanism, favouring lower stellar densities with a longer relaxation timescale, 
allowing the longer survival of the initial substructure and a more extended final distribution.

\subsection{Stellar dynamics}

In this section, we focus on the dynamics of individual stars and study the influence of the star cluster formation scenario.
Our interest is on escaping stars, due to various dynamical interactions in the densest regions of the star cluster. We then study binary stars, as they are the most likely source of escaping stars
during the early phase of the life of the star cluster.
  
\subsubsection{Escaping stars}

Escaping stars are particularly interesting when they are massive: they can travel long distances in the galaxy and eventually explode as supernova (SN) in a location far from their original birthplace, typically in the diffuse ISM. In the kiloparsec scale simulations of \cite{Hennebelle:2014fm} and \cite{Iffrig:2015iu}, the global star formation rate in the Galaxy was reproduced if supernovae were allowed to explode up to 20~pc from their natal cloud, while ``homebred'' supernova explosions were much weaker in suppressing star formation . Similarly, \cite{Gatto15} showed that allowing SN to explode at random positions, rather than at density peaks significantly changes the properties of the ISM, resulting in a hot gas filled volume ISM in the first case and a filamentary, hot gas deprived  ISM in the second one. Thus, being able to predict the correct number of escaping massive stars to be used as input in galactic scale simulations is of vital importance. 

Escaping stars (or for short ``escapers'') are usually categorised into ``walkaway stars'' and ``runaway stars''\footnote{Hypervelocity stars are here considered an extra category,
which is not treated in this work. These stars are thought to have a Galactic centre origin \citep{Brown05}, probably resulting from close encounters between binary systems and the central supermassive black hole. They reach velocities of $\sim$1000 km/s, and hence they are actually unbound from the Galaxy. The current fraction of known hypervelocity stars is $\sim$ 10$^{-8} \% $ of all stars in our Galaxy \citep{Brown07}. }.
Runaway stars (RS) are defined as stars with velocities larger than 30 km/s \citep{Blaauw61}, produced either by supernova explosion in a tight binary system,
during which the companion star of the supernova gets expelled \citep{PZ00, Eldridge11} 
or through dynamical ejection due to very close, three body encounters with massive stars \citep{Perets12, Banerjee12, Oh16}. 
In this section, we focus only on the latter mechanism, while the former can be thought of as a direct consequence of the multiplicity function which we will discuss in the next section.
Walkaway stars (WS , velocities lower than 30 km/s) are normally defined as ``slow escapers", since these are slowly moving stars ejected though normal relaxation processes, 
such as evaporating stars though distant two-body encounters with other single stars or soft binaries \citep{Spitzer87}.

\begin{figure*}
\includegraphics[width=0.85\textwidth]{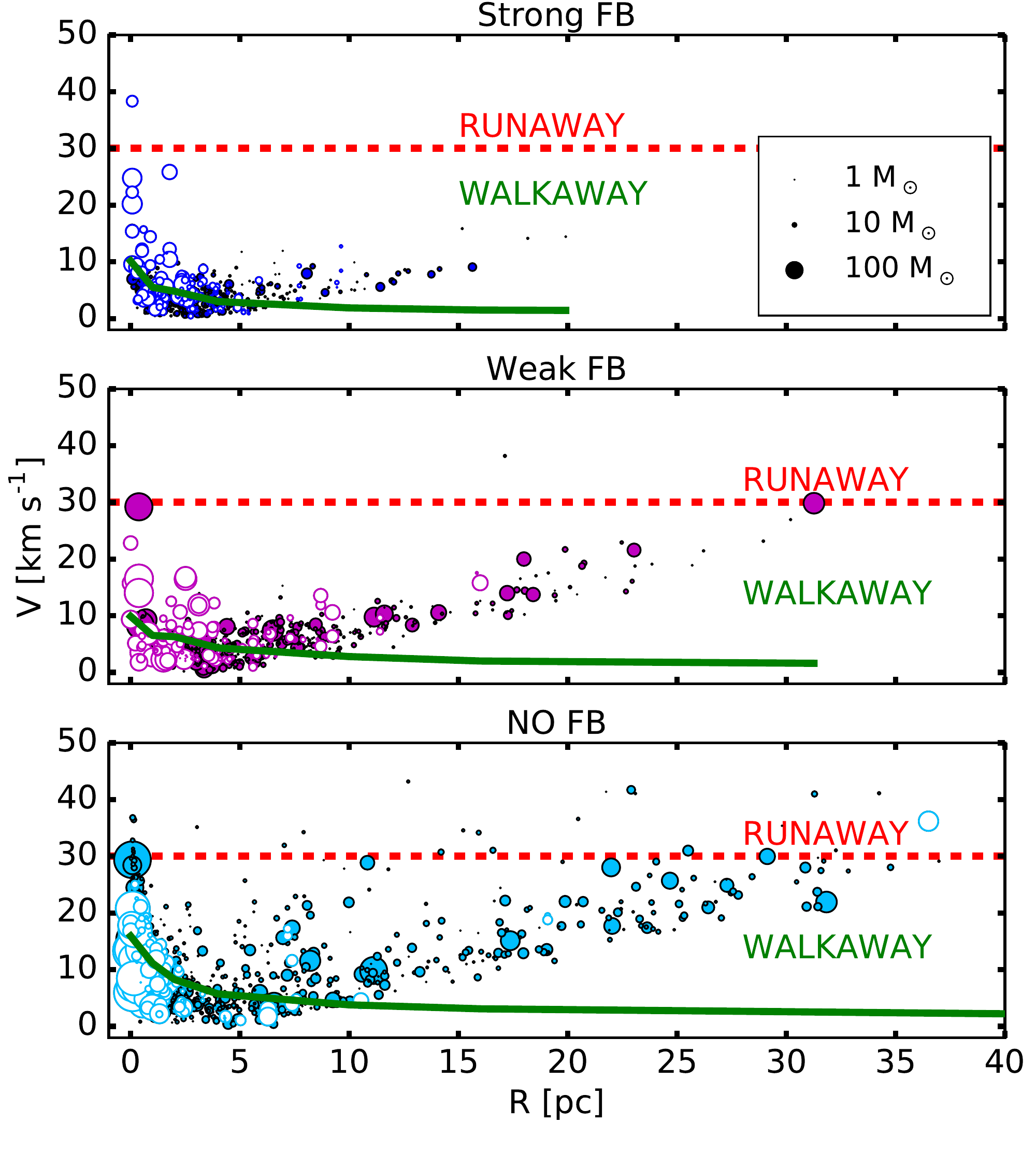}
\caption{Velocity-position diagram of all stars in the cluster for all 3 simulations at t= 2Myr. The symbol size is proportional to the stellar mass. Filled circles: single stars. Open circles: stars which are part of multiple systems (binaries, triple systems etc). The dashed red horizontal line indicates the conventional velocity above which stars are classified as ``runaway''. The solid green line indicates the escape velocity.}
\label{fig:escapers}
\end{figure*}

\begin{table}
\begin{center}
\caption{Statistics about escapers and multiple systems for all simulations (Strong FB, Weak FB, No FB ). In brackets are indicated the percentages, when relevant. }%We list (in order): the number of binary systems, the number of triple systems, the number of 4/5/6 etc -body systems, sum of all multiple systems, total number of stars part of multiple systems, total number of stars with mass > 1 \msun in multiple systems,  total number of stars with mass > 10 \msun in multiple systems, number of bound stars, number of unbound stars, number of runaway stars, number of runaway stars part of a multiple system,  number of walkaway stars, number of walkaway stars part of a multiple system. } %\leavevmode
\label{tab:escapers}
\begin{tabular}{ l | c || c || c }
&Strong & Weak & No FB \\
\hline
\hline
2-body systems 	&51 		&39		&40\\
3-body systems 	& 9		&9		&5\\
>3 body systems 	&6		&5		&5\\ 
Tot multi-body syst 	&66 		&53 		& 50 \\
Stars in multi-body syst           &155(19\%) 		& 126(15\%) 		&150(11\%)\\
	with M>1\msun               & 128(31\%)		&109(20\%)		&129(12\%)\\
	with M>10\msun             & 70(55\%)			&78(39\%)		&94(24\%)\\
\hline  
\hline
Bound stars	&510	 (62\%)	&473	 (57\%)	&900 (64\%)\\
Unbound stars	&315 (38\%)	&355 (43\%)	&502 (36\%)\\
\hline
\hline  
Runaway 			&1 (1\permil)		&3 (4\permil)		&31 (2\%) \\
of which in multi syst	&1				&0				&2\\
Walkaway			&230(28\%)		&297 (36\%)		&476 (34\%)\\
of which in multisyst	&60(7\%)		&50	(6\%)		&39(3\%)\\
\end{tabular}
\end{center}
\footnotesize{}
\end{table}

In Figure~\ref{fig:escapers}, we plot the modulus of velocity versus position of all stars in the cluster. 
The size of the symbols is proportional to the mass of the star. 
Filled symbols indicate single stars, while open circles denote stars which are part of a multiple system (binary, triple or more).
The escape velocity is computed as a function of radius (green solid line in Fig.~\ref{fig:escapers}), 
assuming spherical symmetry, which is a good approximation at t=2Myr (see Fig.~\ref{fig:q}), 
by averaging over the individual escape velocities at different positions within the same spherical shell.

The no feedback case exhibits the highest number of RS candidates\footnote{It is important to clarify that in Figure \ref{fig:escapers} for binaries, triple systems and more,  we plot the true velocity, not the velocity of the centre of mass of the multiple system. Thus, some very high velocity binary members are actually still bound. In the computation of the number of RS we did not correct for this, hence we prefer to talk about RS ``candidates'', meaning that some are probably not unbound yet, but very likely to be, due to frequent interactions with other particles.}, namely 31, or about 2 \% of the total number of stars in the cluster.  
In the cases with feedback, the number of RS is lower, only 1 and 3 in the strong and weak feedback respectively, accounting only for 0.1\% and 0.4\% of the total number of stars. 
The RSs in our simulations are  generally only massive stars (38\msun, 229\msun, 132\msun, 2\msun) in the feedback cases, 
while in the no feedback they cover the whole mass spectrum, going from 0.15 \msun to 417 \msun.
{\reviewtwo The fact that RS are close in mass to the most massive stars in the cluster is easily explained considering the mechanism through which these fast stars formed. Indeed, RS are originated as escaping members of perturbed binary systems, which in our case are mostly composed by massive stars. Due to three body interactions, the lighter member of the binary can escape. RS will therefore have very high masses, close in mass but still lighter than the original massive companion.}

Regarding WS, the fraction changes slightly depending on the exact definition used. 
A first possibility is to take all stars with velocity higher than the escape velocity at a given radius and lower than 30 km/s. 
This gives us a percentage of WS similar in all simulations, around 30\%. 
If we remove stars in multiple systems that are still bound (see Footnote~2), then the fraction is reduced to 20\%.
The final option is to consider WS only in the outskirts of the star cluster, in order to avoid counting stars that are only momentarily unbound. 
If we call R$_{\rm esc}$ the radius at which the escape velocity becomes comparable to the average stellar velocity at that radius, 
we can impose the extra-requirement to be at a distance greater than $R_{\rm esc} \simeq 5$~pc from the centre of the star cluster. 
In this case, we get a very conservative estimate of the fraction around 15\% of the total number of members of the cluster.

Table \ref{tab:escapers} gives an overview of the statistics for escaping stars and multiple systems. 
We also report the fractions of bound and unbound stars, derived by calculating the kinetic and potential energy for every star, 
and then verifying whether the sum of the two energies is negative and positive, respectively.
In all simulations the fraction of bound stars is about the same, around 60\%.
%should we say why unbound doesn't not correspond to WS+RS?

Comparing the populations of RS and WS in the three simulations, we find that the run without feedback produces much more fast escaping stars than the two feedback cases. 
This is consistent with our conclusions in the previous sections, of a very dense star cluster hosting a central clump of tight multiple systems of fast massive stars. 
Three-body interactions can cause the violent ejection of a member of a binary, of the perturber or of the entire binary system (see Fig.~\ref{fig:escapers}).  
In the feedback cases the central densities are lower, and therefore RS stars are rare events. 

The number of WS follow the same trend, with the strong feedback case having slightly less WS stars than the weak and no feedback cases. 
Strong feedback leads to the less frequent interactions, owing to the lower stellar density, which slows down the evaporation of the stars. 
We also notice that the different conditions in the three runs have an effect on the typical velocity and mass of WS.
In the strong feedback case, they don't reach velocities higher than 10 km/s and are mostly low mass stars, probably escaping due to several repeated low energy kicks, 
typical of evaporation, while in the no feedback case both low- and high-mass stars can reach velocities close to the RS limit of 30~km/s, as a result of direct ejection. 

\subsubsection{Multiple systems}

\begin{figure*}
\includegraphics[width=\textwidth]{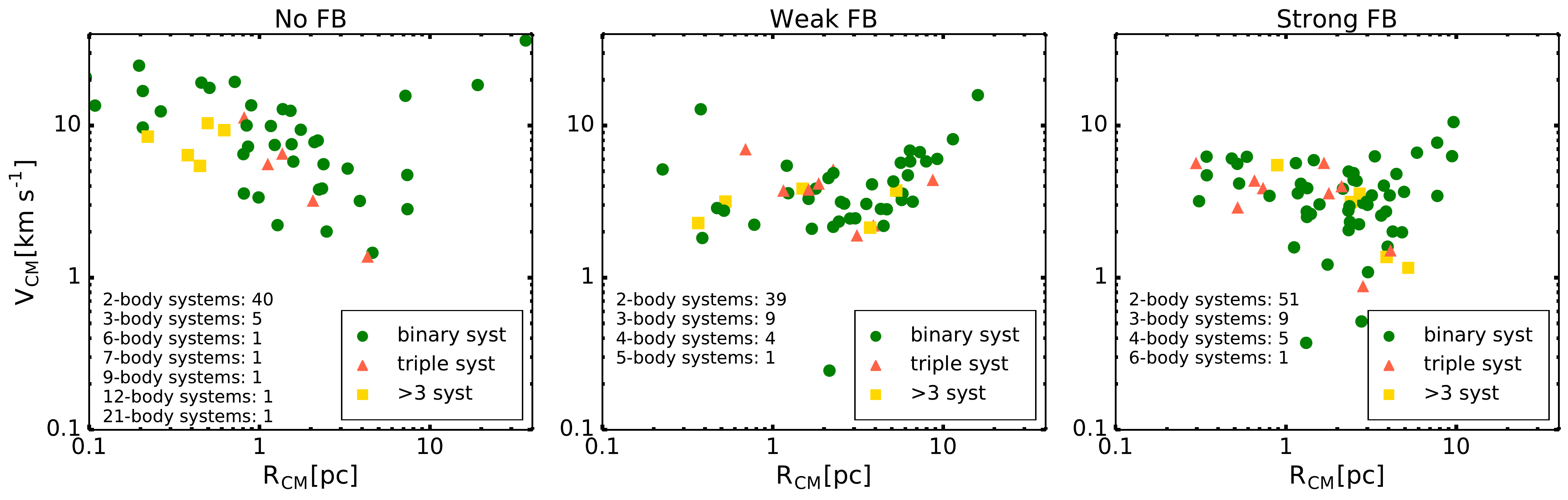}
\caption{Velocity-position diagram of multiple systems (binary, triple, > 3) in the cluster for all 3 simulations. We consider velocities (V$_{CM}$) and positions (R$_{CM}$) of the centres of masses of the multi-systems of stars. In text we give details about the statistics of the systems.}
\label{fig:dist_bin}
\end{figure*}

We focus now on the analysis of multiple stellar systems. 
We identify candidate multiple systems by analysing all possible pairs of stars from the cluster. For each pair we calculate the internal energy, as the total energy of the system in the frame of their centre of mass \citep{BinneyTremaine},
\begin{equation}
\widetilde{E}=\frac{1}{2} \mu {v^2_{12}} - \frac{G m_1 m_2}{r_{12}},
\end{equation}
where $\rm m_1$ and $\rm m_2$ are the masses of the two stars, $\rm \mu=m_1 m_2 / (m_1+m_2) $ is the reduced mass, $\rm v_{12}$ is the relative velocity, $\rm r_{12}$ the relative distance between the two stars, and G is the gravitational constant.

We define the two stars as a binary when $\rm \widetilde E < 0$. We consider all the binary connections as edges in a graph, whose nodes are all the stars involved in multiple systems. 
We use graph reduction algorithms to extract which edges share the same nodes, and we group the nodes together, defining triple, quadruple or quintuple systems in this way.
For example, two binary systems, (i, j) and (j, k), which share one node, are considered a triple system. 

A slightly different technique was used by \cite{Bate2009} to identify multiple systems. 
They replaced the binary systems by a virtual star sitting at the centre of mass and with mass equal to the sum of the two masses. 
They then searched for isolated stars with a negative binding energy with these virtual stars. 
The same procedure was iterated only up to quadruple systems. 

An advantage of our graph-based method is that we can easily identify systems with multiplicity larger than 4. 
However, in most cases the two algorithms will produce the same catalogue of multiple systems, 
since, in our case, most multiple systems include a massive star, which dominates the gravitational potential of the system (see Fig.~\ref{fig:mass_bin}).

In Table~\ref{tab:escapers}, we report on the statistics of binary, triple and more than 3-body systems for all three simulations. 
We note that the fractions of stars in multiple systems, also known as the multiplicity fraction, correlates with the strength of feedback, 
with overall percentages spanning from 11\% (no feedback) to 19\% (strong feedback). 
If we exclude stars with mass lower than 1~\msun, 
the multiplicity fraction differentiates even more between the three feedback regimes and rises to 12\%, 20\% and 31\% for no, weak and strong feedback respectively.
For stars, with mass greater than 10~\msun, the fraction goes up to 24\%, 39\% and 55\%. 
{\review Due to the adopted sink density threshold, fragmentation is not fully resolved for low-mass stars, which might contribute to lower the multiplicity fraction of low-mass stars. A more detailed study focused on the multiplicity of low-mass pre-stellar cores was performed by \cite{Lomax2015}.}

The observed multiplicity fraction is around 20\%, when one considers field stars and low mass stars, but reaches 60\% for OB and massive stars \citep{Kraus07, Lafreniere08, Goodwin10}. 
These values are well reproduced by our strong feedback case, while our no feedback run underestimates the number of stars in multiple system, when compared to observations, especially for massive stars. 
Observations also reveal that the binary fraction is higher in lower density star forming regions, like in our strong feedback case,
while denser clusters exhibit multiplicity fractions comparable to the field or low mass stars, like in our no feedback case \citep{Reipurth07, Lafreniere08}.

In Figure~\ref{fig:dist_bin}, we plot the distribution of multiple systems in terms of position versus velocity. 
Here, we consider the positions and the velocities of the centres of masses, explaining why velocities are lower than in Figure~\ref{fig:escapers}. 
In general, we observe that in the feedback simulations binary, triple and more than 3-body systems are uniformly distributed throughout  the cluster, 
while the no feedback  case shows many systems with very high multiplicity in the very inner part of the cluster, while binaries and triple systems occupy the outskirts. 
In all cases, we see many ejections of binary systems. 

In the same plot, we also indicate the exact count of multiple systems, in particular for groups with more than 3 bodies. 
We notice that the maximum multiplicity reaches a much higher value in absence of feedback, due to the very high stellar density.
With feedback, the most crowded multiple systems have 5 or 6 members, while in the no feedback run we have systems with as many as 9, 12 and 21 members. 
All these high multiplicity systems are highly unstable and they will be destroyed during the subsequent dynamical evolution of the cluster.
As a matter of fact, we do not observe such systems in real star clusters . 

\begin{figure*}
\includegraphics[width=\textwidth]{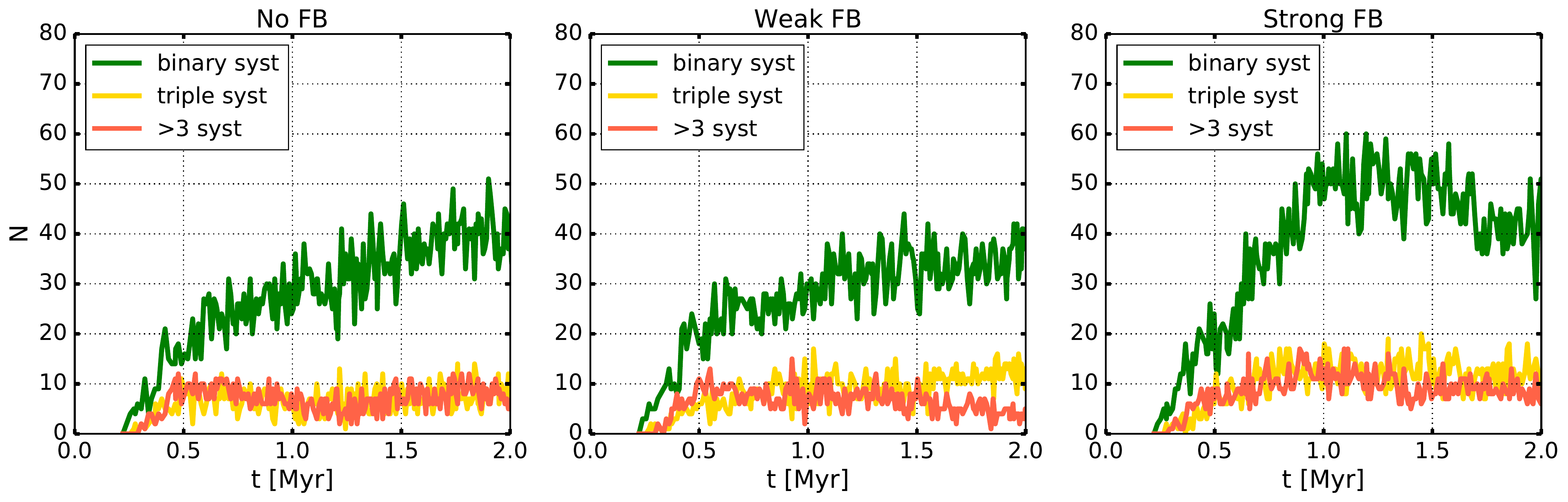}
\caption{Evolution with time of the number of multiple systems (binary, triple, > 3).}
\label{fig:bin_t}
\end{figure*}

In the strong feedback case, the lower stellar density will also guarantee the survival of the binary systems, 
which otherwise, like in the no feedback case, aggregate in bigger associations or are destroyed in three-body interactions.
In that context, it is useful to divide binaries into two categories, {\it soft} binaries and {\it hard} binaries. 
Soft binaries are systems for which $ |\widetilde E|< {\overline K}$, while hard binaries have $|\widetilde E|>{\overline K}$, 
where ${\overline K}$ is the typical kinetic energy of the stars in the cluster \citep{BinneyTremaine}. We use here the median kinetic energy.
According to this definition, for the two feedback cases, we have 50\% hard and 50\% soft binaries,  
while the no feedback case shows only 30\% hard and 70\% soft binaries, which support even more our conclusion that binaries will survive longer in the strong feedback case.

In Figure~\ref{fig:bin_t}, we plot the time evolution of the number of binary, triple and more than 3-body systems. 
In all three models, the number of triple (or more) systems is almost constant. This is not the case for the number of binaries. 
In the strong feedback case, it increases sharply during cloud collapse and after the gas has been dispersed around 1~Myr, it slowly decreases.
No additional stars are created and the soft binaries get destroyed through ejection or evaporation. 
In the no feedback case, the number of binaries keeps increasing since star formation continues until the end of the simulation. 
The weak feedback case shows an intermediate behaviour, with a mild initial increase, 
followed by a almost constant evolution.

\begin{figure*}
\includegraphics[width=\textwidth]{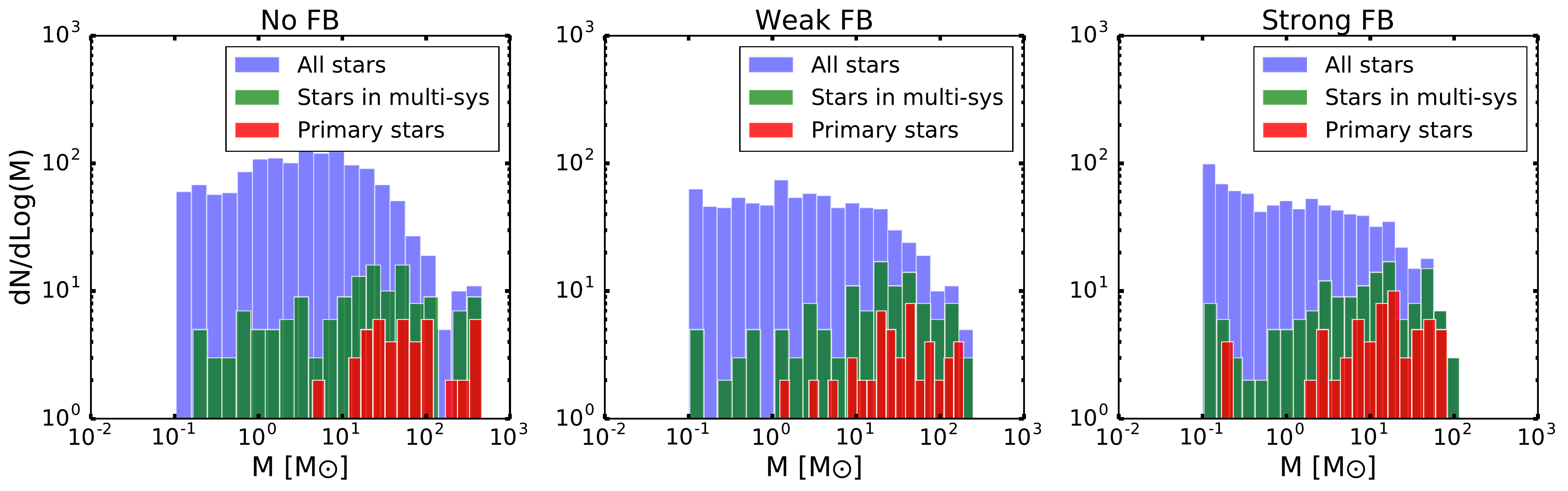}\;
\caption{Mass diagrams of stars in multiple systems . The blue histogram reports the total mass function of the cluster. In green we show only stars which are member of multiple systems (binaries, triple, >3) and in red we plot the mass of the primary star, intended as the most massive star in the multi-body system. Most multiple systems contain at least one star with M > 10\msun. In the case with feedback we have also some systems composed only from low-mass stars. }
\label{fig:mass_bin}
\end{figure*}

\section{Summary and Discussion}
\label{discussions}

In this work, we have performed radiation hydrodynamics simulations of a collapsing turbulent molecular cloud with the adaptive mesh refinement code {\ttfamily RAMSES}.
We have studied in detail the emergence of the star cluster from the parent gas cloud with and without the influence of photoionisation feedback. Stars are modelled using a sink particle algorithm. 
Photo-ionising radiation is included with two different regimes: weak and strong feedback. 
We also perform a reference simulation without any feedback. 
Our main focus is the emerging properties of the star cluster, both from a structural and a dynamical point of view.

The main effect of photo-ionisation feedback is to reduce considerably the stellar density of the star cluster and to limit the accretion on very massive stars. 
This has a large impact on the dynamical properties of the final star cluster. 
As a result of the reduced stellar density, the star cluster can settle in virial (or even sub-virial) state, 
while in the absence of feedback, strong and frequent close interactions in a highly collisional environment lead to the disruption of the cluster.
This is in contrast with the traditional view that strong feedback is responsible for the star cluster early mortality, by rapidly removing gas from the emerging cluster \citep{Hills80}. 
The star formation efficiency can be reduced down to 20\%, without affecting the virial equilibrium of the star cluster. 

The stellar mass function is also affected at the high mass end, with a clear self-regulating role played by feedback, 
limiting the mass of the most massive stars by a factor of 4 compared to the no feedback case. 
As a result, our mass function with strong feedback compares favourably with observations of two starburst clusters (NGC~3603 and Arches) 
but only after re-normalising the data and for masses larger than $10 M_\odot$. 

We also use mass segregation statistics to test our model. In absence of feedback, the higher stellar density causes an unrealistically too high degree of mass segregation for a few very massive stars.
When including strong feedback, we obtain a more extended star cluster with a degree of mass segregation consistent with the one observed in NGC~3603. 

We have also computed the number of ejected stars, which anti-correlates with the feedback strength:
for weaker feedback, we get a higher stellar density and more stars are escaping, both as runaway and walkaway stars. 
This result has profound implications for galactic evolution, when supernovae will start exploding at later time in a large variety of galactic environments. 

Our statistics of multiple systems of stars supports the same conclusion: in a denser environment, the fraction of stable binary systems is lower, 
and most stars tend to either cluster into unstable many-body systems, or are ejected. 
On the other hand, in the strong feedback case, the lower stellar density guarantees the survival of a higher fraction of binaries, in better agreement with observations.

Our results are in line with the findings of \cite{Dale15,Dale12a, Dale13a}, which showed that photo-ionisation feedback effectively lowers the star formation efficiency,
and, for low-mass clouds like ours, can expel most of the gas within 3~Myr, before the first supernova can explode. 
\cite{Parker13, Parker15, Parker15b} also observed that photo-ionisation feedback reduces the stellar density in the emerging cluster,
which allows substructures to survive longer than in a scenario without feedback. 
However, in contrast with \cite{Parker15} who did not find any mass segregation in the feedback case, we do see a weak mass segregation signature, which is well in agreement with observations.
Interestingly, although \cite{Fujii15b} found that a local star formation efficiency of at least  50\% is necessary for the formation of young massive clusters, 
we could reach a value as low as 20\% and still form a bound star cluster.  

Our goal in this work is to better understand the transition from a gas cloud to a stellar cluster, or in other words, from gas dynamics to stellar dynamics.
In that context, our direct N-body integrator, a second order leap frog scheme is probably accurate enough for our relatively short time integration, 
but its accuracy is far below the required standards in stellar dynamics for longer time scales. 
This sets the limit on the runtime of our simulations to a couple of Myr. 
This explains why, in comparison to \cite{Parker15}, who were able to investigate the long term evolution of the star cluster, we are forced to limit our study to the first 2~Myr. 

We have also decided in this work to focus exclusively on photo-ionisation radiation. 
We have therefore neglected magnetic fields and other radiation processes, but also other important physical processes that could be relevant.
Supernovae explosions, for example, are ignored, but, given the cloud mass we have adopted, all the gas is removed from the star cluster after only 2~Myr and they are therefore irrelevant.
For larger cloud masses, however, this would not be the case. We have also ignored the possible role of stellar winds, 
but these have been shown to be negligible compared to photo-ionisation feedback \citep{Dale15}.

We have also ignored the effect of the UV radiation force (or UV radiation pressure) on the gas dynamics.   
It has been shown that momentum transfer from photo-absorptions is only relevant for ultra compact HII regions, with densities larger than $10^{-15}$~g/cc 
and sizes smaller than $10^{-3}$~pc, completely unresolved in our simulations \citep[see e.g.][]{Rosdahl2015}. 
More relevant would be the inclusion of lower energy photons, in the optical and infrared range. 
These propagate from accreting stars through dust  grains, and are scattered into new infrared photons.
Inside the HII regions we can probably ignore these effects as dust is quickly sublimated at $10^4$ K, however, infrared and optical radiation can play a role before massive stars form.
\cite{Skinner2015} have shown that infrared radiation has very little impact on the gas removal and on the cloud destruction for realistic values of the dust opacity.
Infrared radiation is likely to play a more important role on the fragmentation of molecular cores, but at scales we also do not resolve in our simulations.
 
In summary, we are able to simulate the collapse of a molecular cloud and the emergence of a star cluster, whose properties are tightly connected to the gas dispersal process. 
Comparing our results to two observed, very young and still active star cluster, NGC~3603 and Arches,
we conclude that an initially sub-virial molecular cloud with a star formation efficiency lower than 30\% can reproduce observations fairly well. 
Our analysis provides useful insights also for simulations on galactic scales. 
Star clusters are indeed the building blocks of galaxy formation and evolution.  
Understanding in details  their properties, such as mass segregation, mass and multiplicity functions and escaping stars statistics, just after they emerged from their parent cloud,
is of primary importance for their longer term dynamical evolution, but also for the evolution of their host galaxies.

%\section{Conclusions}
%
%The last numbered section should briefly summarise what has been done, and describe
%the final conclusions which the authors draw from their work.
%
\section*{Acknowledgements}
We thank the anonymous referee for their thoughtful comments and helpful suggestions.
This work is supported by the STARFORM Sinergia Project funded by the Swiss National Science Foundation. 
We would like to thank Sam Geen, Patrick Hennebelle and Michela Mapelli for useful discussions. 

%%%%%%%%%%%%%%%%%%%%%%%%%%%%%%%%%%%%%%%%%%%%%%%%%%



\bibliographystyle{mnras}
\bibliography{biblio,romain} % if your bibtex file is called example.bib

%%%%%%%%%%%%%%%%%%%%%%%%%%%%%%%%%%%%%%%%%%%%%%%%%%

%%%%%%%%%%%%%%%%% APPENDICES %%%%%%%%%%%%%%%%%%%%%

%this figures should be done better

%%%%%%%%%%%%%%%%%%%%%%%%%%%%%%%%%%%%%%%%%%%%%%%%%%

% Don't change these lines
\bsp	% typesetting comment
\label{lastpage}
\end{document}